\documentclass[reprint, nofootinbib, prb, superscriptaddress]{revtex4-1}
\usepackage{amsmath} \usepackage{graphicx} \usepackage{dcolumn}
\usepackage{bm} 
\usepackage{amssymb}
\usepackage{pstricks}

\begin{document}

\newlength{\figurewidth}
\setlength{\figurewidth}{\columnwidth}

\newcommand{\prtl}{\partial}
\newcommand{\la}{\left\langle}
\newcommand{\ra}{\right\rangle}
\newcommand{\dla}{\la \! \! \! \la}
\newcommand{\dra}{\ra \! \! \! \ra}
\newcommand{\we}{\widetilde}
\newcommand{\smfp}{{\mbox{\scriptsize mfp}}}
\newcommand{\smp}{{\mbox{\scriptsize mp}}}
\newcommand{\sph}{{\mbox{\scriptsize ph}}}
\newcommand{\sinhom}{{\mbox{\scriptsize inhom}}}
\newcommand{\sneigh}{{\mbox{\scriptsize neigh}}}
\newcommand{\srlxn}{{\mbox{\scriptsize rlxn}}}
\newcommand{\svibr}{{\mbox{\scriptsize vibr}}}
\newcommand{\smicro}{{\mbox{\scriptsize micro}}}
\newcommand{\scoll}{{\mbox{\scriptsize coll}}}
\newcommand{\sattr}{{\mbox{\scriptsize attr}}}
\newcommand{\sth}{{\mbox{\scriptsize th}}}
\newcommand{\sauto}{{\mbox{\scriptsize auto}}}
\newcommand{\seq}{{\mbox{\scriptsize eq}}}
\newcommand{\teq}{{\mbox{\tiny eq}}}
\newcommand{\sinn}{{\mbox{\scriptsize in}}}
\newcommand{\suni}{{\mbox{\scriptsize uni}}}
\newcommand{\tin}{{\mbox{\tiny in}}}
\newcommand{\scr}{{\mbox{\scriptsize cr}}}
\newcommand{\tstring}{{\mbox{\tiny string}}}
\newcommand{\sperc}{{\mbox{\scriptsize perc}}}
\newcommand{\tperc}{{\mbox{\tiny perc}}}
\newcommand{\sstring}{{\mbox{\scriptsize string}}}
\newcommand{\stheor}{{\mbox{\scriptsize theor}}}
\newcommand{\sGS}{{\mbox{\scriptsize GS}}}
\newcommand{\sBP}{{\mbox{\scriptsize BP}}}
\newcommand{\sNMT}{{\mbox{\scriptsize NMT}}}
\newcommand{\sbulk}{{\mbox{\scriptsize bulk}}}
\newcommand{\tbulk}{{\mbox{\tiny bulk}}}
\newcommand{\sXtal}{{\mbox{\scriptsize Xtal}}}
\newcommand{\sliq}{{\text{\tiny liq}}}

\newcommand{\smin}{\text{min}}
\newcommand{\smax}{\text{max}}

\newcommand{\saX}{\text{\tiny aX}}
\newcommand{\slaX}{\text{l,{\tiny aX}}}

\newcommand{\svap}{{\mbox{\scriptsize vap}}}
\newcommand{\sjam}{J}
\newcommand{\Tm}{T_m}
\newcommand{\sTS}{{\mbox{\scriptsize TS}}}
\newcommand{\sDW}{{\mbox{\tiny DW}}}
\newcommand{\cN}{{\cal N}}
\newcommand{\cB}{{\cal B}}
\newcommand{\br}{\bm r}
\newcommand{\be}{\bm e}
\newcommand{\cH}{{\cal H}}
\newcommand{\cHlt}{\cH_{\mbox{\scriptsize lat}}}
\newcommand{\sthermo}{{\mbox{\scriptsize thermo}}}

\newcommand{\bu}{\bm u}
\newcommand{\bk}{\bm k}
\newcommand{\bX}{\bm X}
\newcommand{\bY}{\bm Y}
\newcommand{\bA}{\bm A}
\newcommand{\bb}{\bm b}

\newcommand{\lintf}{l_\text{intf}}

\newcommand{\DV}{\delta V_{12}}
\newcommand{\sout}{{\mbox{\scriptsize out}}}
\newcommand{\dv}{\Delta v_{1 \infty}}
\newcommand{\dvin}{\Delta v_{2 \infty}}

\newcommand{\wtp}{\tilde{p}}
\newcommand{\wtK}{\widetilde{K}}
\newcommand{\wtgm}{\tilde{\gamma}}
\newcommand{\wtg}{\widetilde{g}}

\def\Xint#1{\mathchoice
   {\XXint\displaystyle\textstyle{#1}}%
   {\XXint\textstyle\scriptstyle{#1}}%
   {\XXint\scriptstyle\scriptscriptstyle{#1}}%
   {\XXint\scriptscriptstyle\scriptscriptstyle{#1}}%
   \!\int}
\def\XXint#1#2#3{{\setbox0=\hbox{$#1{#2#3}{\int}$}
     \vcenter{\hbox{$#2#3$}}\kern-.5\wd0}}
\def\ddashint{\Xint=}
\def\dashint{\Xint-}
\title{Low-temperature anomalies in disordered solids: A cold case of
  contested relics?}

\author{Vassiliy Lubchenko} \email{vas@uh.edu}
\affiliation{Departments of Chemistry and Physics, University of
  Houston, Houston, TX 77204}

\date{\today}

\begin{abstract}

  Amorphous solids manifest puzzling effects of mysterious degrees of
  freedom that give rise to a heat capacity and phonon scattering in
  great excess over what would be expected for a solid that has a
  unique vibrational ground state. Of particular conceptual importance
  is the apparent near universality of phonon scattering in amorphous
  solids made by quenching a liquid.  To rationalise this
  universality, scale-free scenarios have been proposed that either
  hinge on there being long-range interactions between bare structural
  degrees of freedom or that invoke long-range criticality stemming
  from the emergence of marginally stable vibrational modes. In a
  contrasting, local scenario, the puzzling low-temperature degrees of
  freedom are, instead, weakly-interacting, strongly anharmonic
  degrees of freedom each of which involves the motion of a few
  hundred particles.  In this scenario, the universality of phonon
  scattering comes about because the characteristic energy scale of
  the local anharmonic resonances and the strength of their
  interaction with phonons are both set by the glass transition
  temperature $T_g$, while their concentration is set by the
  cooperativity size $\xi$ for dynamics at $T_g$. The nanoscopic
  length $\xi$ is manifested in vibrational excitations of the spatial
  boundary of the resonances, which underlie the so-called Boson peak,
  and very deep, topological midgap electronic states in glassy
  semiconductors, which are implicated in a number of strange
  optoelectronic phenomena in amorphous chalcogenides. I discuss the
  merits of the above scenarios when confronted with experimental
  data. \\


\end{abstract}

\maketitle


\section{Introduction}


The detailed mechanism of the emergence of mechanical rigidity in
structural glasses remains a disputed topic, enough so as to have
found its way into the popular culture.~\cite{NYTGlasses} The present
article focuses on a specific subset of structural degrees of freedom
in frozen glasses whose microscopic origin has attracted much
attention since the early 1970s~\cite{ZellerPohl, AHV, Phillips,
  LowTProp, Pohl_review, Esquinazi} and is at the heart of that
debate.  These puzzling degrees of freedom would not be present in
solids that have a unique vibrational ground state and thus imply that
structural glasses are, in fact, vastly
structurally-degenerate. Experimental signatures of these degrees of
freedom are seen down to the lowest temperatures accessed so far in
experiment---i.e. fractions of the Kelvin---thus suggesting the
spectrum of the alternative structural states is truly gapless while
the barriers separating such states can be arbitrarily low.

To begin discussing mechanical instabilities in low-temperature
solids, we set the stage by invoking Nernst's Law and symmetry
considerations. According to the Nernst theorem, any dynamical degree
of freedom, when equilibrated, must attain a unique quantum state as
the temperature is lowered to the absolute zero; otherwise it must be
supplanted by other degrees of freedom, for instance, by way of a
phase transition. In addition to being relatively scarce, one may
reasonably expect low energy states to be also relatively symmetric:
On the one hand, degenerate eigenstates of the Hamiltonian comprise an
irreducible representation of a symmetry group.~\cite{LLquantum} On
the other hand, lower energy solutions of the Schr\"odinger equation
have fewer nodes implying a progressively smaller size of the
corresponding irreducible representation.  Thus the ground state is
expected to be described by the totally symmetric
representation. Higher energy states typically will be increasingly
more degenerate and, at the same time, less symmetric.

When particles combine to form a solid, translational symmetry is
drastically lowered while the (quantum-mechanical) kinetic energy is
increased. A solid can not be the true ground state of the system in
the absence of external field: Quantum-mechanical superpositions of
degenerate symmetry-broken states are clearly lower in energy. Yet for
practical situations, the tunnelling barrier that would allow the
system to escape such a symmetry-broken, ``classical'' state is
incomprehensibly high.~\cite{AndersonBasicNotions} While not being
true ground states, the structures of actual solids formed on long
time scales still represent low energy, relatively unique
configurations. Uniqueness is important in the present context because
it tautologically implies stability. Thus, solids produced as a result
of very slow cooling may by expected, quite reasonably, to be both
stable on every length scale and exhibit a great deal of spatial
symmetry.  In contrast, solids made through routes other than slow
equilibration of a melt can not be generally expected to adopt such
low energy configurations or be stable at each wavelength, even if
exhibiting substantial stability macroscopically. Nor should such
hastily-made solids be expected to be periodic, implying a lack of
point symmetries.  Technologically important examples of strongly
off-equilibrium solids are quenched glasses, deposited films, and
rubbers among many others.

The present article focuses primarily on those aperiodic solids made
by quenching a liquid sufficiently rapidly---via cooling or
compression---so that the liquid fails to crystallise. Instead, the
particles become kinetically arrested in a structure that is
approximately a snapshot of a liquid. In reference to window glass,
which is made by cooling a silicate melt, all solids made by quenching
a liquid are {\em also} called glasses, whether transparent or
not. Judging by their strength, glasses are often rigid---despite
their being aperiodic---in some sense more rigid than many periodic
crystals. It then comes as a surprise that glasses apparently host
structural degrees of freedom whose density of states is considerably
greater---at pertinent energies---than the density of states of the
vibrational modes of an elastic medium with a unique vibrational
ground state.

The discovery of these fascinating degrees of freedom came about in a
rather prosaic way:~\cite{ACAnderson_pc} At least as early as the
1960s, it became clear that the ordinary adhesive epoxy, which is
often used to affix samples in various cryogenic experiments, conducts
heat much less robustly than does a crystalline solid.  At
sufficiently low temperatures, the mean free path of a phonon in most
crystals is so large that it exceeds the dimensions of the typical
sample. Under these circumstances, the heat {\em conductance} of the
sample does not scale inversely with the sample length but is simply
determined by the specific heat of the phonons, which follows the
Debye $T^3$ law.~\cite{Kittel} On the relevant lengthscale, one cannot
even define the heat conductivity, it being a bulk property. In
contrast, epoxy---which is a glassy substance---exhibits phonon mean
free paths that are much shorter than the sample dimensions down to
very low temperatures so that its heat {\em conductivity} remains
perfectly well defined. The measured heat conductivity of glasses
scales nearly quadratically with temperature, viz., $T^{2-\alpha'}$,
$\alpha' = 0.05 \ldots 0.2$.  The latter observation became recognised
relatively widely after the seminal work of Zeller and
Pohl~\cite{ZellerPohl} who crucially established also that the
apparent excess phonon scattering in glassy solids is accompanied by a
heat capacity in excess of the Debye $T^3$ law. This excess heat
capacity scales nearly linearly with temperature, $T^{1+\alpha}$,
$\alpha = 0.1 \ldots 0.3$.

The presence of such a large number of residual degrees of freedom
then called for a revision of the conventional notions of mechanical
stability of solids based on a unique ground state. A simple scenario
was conceived shortly thereafter, by which the stability is
compromised but only locally and quite sporadically: Owing to the
ostensive lack of local point-symmetries on atomic lengthscales in
amorphous solids, one might expect that the lattice might not be
stable locally everywhere. At the same time, an amorphous solid can be
regarded as isotropic on large lengthscales, the better the larger the
length. One can put these two notions together by imagining small,
bi-stable or, more generally, multi-stable groups of atoms that are
embedded in a macroscopically stable lattice. Such a multi-stable
group of atoms represents a multi-level system, but only the two
lowest energy states will be relevant at sufficiently low
temperatures. Thus one arrives at the venerable two-level system (TLS)
model, which is often referred to as the standard tunnelling model
(STM) when the distribution of the off-diagonal matrix element
$\Delta$ is explicitly specified.~\cite{AHV, Phillips, Jackle,
  LowTProp} Indeed, one may make a reasonable assumption that the
distributions for the on-site energy splitting $\epsilon$ and
tunnelling barrier $V^\ddagger$ each vary slowly as both quantities
approach zero, as pertinent to low temperatures. One consequently
obtains a unified, phenomenological description for several cryogenic
anomalies observed in glasses that employs only a few adjustable
parameters characterizing the number of states and the distributions
of $\epsilon$ and $V^\ddagger$.~\cite{LowTProp} The simplest rendition
of the model is a two-level system coupled to a local dimensionless
strain $\nabla \phi$:
\begin{equation} \label{HTLS} \cH = \frac{\epsilon}{2} \sigma_z +
  \frac{\Delta}{2} \sigma_x + \sigma_z {\bm g} \nabla \phi,
\end{equation}
where $\sigma_z$ and $\sigma_x$ are the usual Pauli
matrices.~\cite{LLquantum} The energy density corresponding to the
(single-component) deformation $\phi$ of the lattice relative to some
reference state is given in the usual fashion by $K (\nabla
\phi)^2/2$, where $K$ is the elastic modulus. Internal degrees of
freedom cannot couple to the absolute displacement $\phi$, hence the
gradient coupling in Eq.~(\ref{HTLS}).  Note that for the simplified,
``scalar'' elasticity from Eq.~(\ref{HTLS}), there is no difference
between longitudinal and transverse sound waves.  We note that the
tunnelling amplitude $\Delta$ is semi-classically computed in the
simplest assumption that the instanton's motion near the tunnelling
bottleneck is harmonic with the under-barrier frequency given by
$\omega^\ddagger$. Thus $\Delta$ depends exponentially on the barrier
height: $\Delta = \Delta_0 e^{-\pi V^\ddagger/\hbar \omega}$. We note
that the minimalistic two-level model can be generalised by
postulating a variety of soft potentials that could in principle
govern the motions of small groups of atoms.~\cite{soft_pot,
  soft_potBuchenau, Klinger} The greater flexibility of such models
naturally allows for good fits of thermal data but at the expense of
using additional adjustable constants.

The apparent two-level nature of the excess degrees of freedom was
directly confirmed shortly after the model was proposed, via
phonon-echo experiments,~\cite{GG} and, much later, in single-molecule
experiments that allow one to monitor the frequency of a local
chromophore embedded in a glassy matrix.~\cite{Orrit, Maier} Quite
tellingly, however, the latter experiments also show a deviation from
the standard model. They show that quite often, a fluorescent molecule
couples to what seems to be a pair TLS at a time or a structural
resonance that has more than two levels.

In the local description, which effectively amounts to placing
localised defects within a stable matrix, the concept of a (thermal)
phonon seems to be still reasonably valid despite the lack of
periodicity: The measured mean free path $l_\smfp$ of a vibrational
wave-packet turns out to be about hundred times longer than its
wavelength $\lambda$; one can still define the frequency and
wavelength sufficiently well.  The macroscopic stability of the solid
is thus handily ensured, too.  Since the $l_\smfp/\lambda$ ratio gives
the number of whole periods, at the pertinent frequency, that fit
within the vibrational wave-packet, it can be thought of as a measure
of how well a phonon is defined as a quasi-particle, a quality-factor
of sorts. For crystals, the phonon quality factor near absolute zero
can be made arbitrarily large thus making the phonons rigorously
defined excitations in the Landau sense.

Serious doubts on the local picture described above were however cast
following the work of Freeman and Anderson,~\cite{FreemanAnderson} who
observed that the quality factor $l_\smfp/\lambda$ ratio varies
surprisingly little over several families of amorphous compounds; the
numerical value of the ratio is close to 150. At the same time, the
$l_\smfp/\lambda$ ratio can be shown within the two-level
phenomenology to be determined by the TLS density of states $\bar{P}$
and the strength $g$ of the TLS-phonon coupling:~\cite{AHV, Phillips,
  LowTProp} $l_\smfp/\lambda \simeq (\bar{P} g^2/K)^{-1}$. Thus one
obtains that in the TLS regime, the following nearly universal
relation holds:
\begin{equation} \label{Uratio} \frac{l_\smfp}{\lambda} \simeq
  \left(\frac{\bar{P} g^2}{K} \right)^{-1} \simeq 10^2.
\end{equation}
Similar conclusions were reached when analysing internal friction
data,~\cite{Leggett} see also Refs.~\onlinecite{BerretMeissner,
  ph_loc}.  The above equation suggests that there is an apparent,
intrinsic relationship between the material constants $g$, $K$, and
$\bar{P}$. Yu and Leggett~\cite{YuLeggett} stressed that this
empirical relation hardly seems a coincidence since it holds across a
large set of substances exhibiting very different local chemistries
which one would think determine local excitions.

Although the numerical value in the r.~h.~s. of Eq.~(\ref{Uratio})
does not seem to be invariant over {\em all} known types of glassy
substances, such as polymers or amorphous metallic alloys, one may
still reasonably ask~\cite{YuLeggett, Leggett} how much local
chemistry matters. One may even ask whether there is a way to
discriminate between detailed microscopic scenarios of the origin of
the excess states.~\cite{doi:10.1021/jp402222g} Is it perhaps the case
that the apparent low-$T$ states in glasses are an emergent property
accompanied by a criticality-induced scale invariance? A similar
notion arises from a seemingly very different type of analysis,
namely, the study of meanfield disordered systems in high
dimensions.~\cite{Berthier26072016} When quenched sufficiently deeply,
such systems exhibit a continuous replica-symmetry breaking (RSB), the
so-called Gardner transition,~\cite{Gardner1985747} that would
naturally supply a plentitude of marginally stable modes,
infinite-range correlations, and, perhaps, scale invariance in three
dimensions.

Another view, espoused by Wolynes and myself,~\cite{LW, LW_BP, LW_RMP,
  LSWdipole} is rooted in the classical density functional theory of
the glassy state itself and its origin in quenching the liquid. This
picture ties the two-level systems to the molecules and their motions
directly in the actual three-dimensional space. In this treatment, the
rigidity of glasses emerges as a result of a translational symmetry
breaking in the form of a random first order transition
(RFOT).~\cite{LW_ARPC, L_AP} In a surprisingly indirect yet natural
way, the near universality in phonon scattering comes about, within
this finite-dimensional framework, because of the way glasses are
made. The relation between the parameters of the two-level excitations
arise from experimental constraints on the rate of quenching, on the
one hand, and an intrinsic connection between the cohesive energy of
the solid and the interaction strength of the phonons with local
degrees of freedom, on the other hand. This intrinsic connection stems
from the marginal stability of small-scale motions against vibrational
excitations of the lattice when the rigidity sets it.  The quantum
states are relatively local and turn out to be only weakly interacting
at low temperatures, much as in the phenomenological tunnelling
model. In contrast with that model, however, the density of states of
the local degrees of freedom is no longer an adjustable parameter but,
instead, is rigidly prescribed by the cooperativity size for activated
transport near the glass transition temperature $T_g$ and the
temperature $T_g$ itself. Interactions between these emergent local
degrees of freedom provide only a weak perturbation on the picture but
still reveal themselves in subtle ways, such as in the otherwise
strange negative thermal expansivity of low temperature
glasses,~\cite{LW_RMP} a conceptually-important yet relatively small
effect.

The three scenarios listed above are quite distinct in both style and
substance.  In the Yu-Leggett scenario, the mechanical stability of
the overall matrix is simply assumed at the onset.  The effective
degrees of freedom are complex composites of local degrees of freedom
interacting so strongly as to completely loose their individuality. In
meanfield RSB analyses, the nature of the excitations is not entirely
clear either; overall rigidity of the matrix arises
internally-consistently in the treatment since the kinetic barriers
separating distinct aperiodic free energy minima are strictly infinite
and so is the correlation length.  The only explicit length scales in
the problem are the molecular length and the vibrational
amplitude. The role of finite-dimensional effects, such as
fluctuation-induced lowering of the symmetry breaking transition is
not obvious. At the same time, perturbative analyses suggest such
effects could completely destroy the transition in three
dimensions.~\cite{AngeliniBiroli}

In the RFOT-based approach, in contrast, there is a finite---and a
substantial indeed---temperature range, in which the barriers are
finite thus allowing for reconfigurations of {\em finite}-sized
regions, the cooperativity length $\xi$ not exceeding a few
nanometers. The system is mechanically stable on lengths less than
$\xi$ but is metastable on larger lengthscales.  The scenario
explicitly describes how the vast majority of strongly-interacting
molecular translational degrees of freedom in the uniform liquid
become frozen out with lowering the temperature so that the remaining
degrees of freedom can be regarded as relatively local and only
weakly-coupled.  Along with the emergence of a finite length scale
$\xi$ there come about very special degrees of freedom that are simply
not accessible to meanfield treatments: Vibrations of the boundary of
a reconfiguring region reveal themselves as an additional,
vibrational-like contribution to the total density of states widely
known as the Boson peak.~\cite{LW_BP, LW_RMP} The latter is an
umbrella term referring to degrees of freedom seen by vibrational
spectroscopies at frequencies near 1 THz and has been also associated
with the ``bump'' in the heat capacity and the so called plateau in
the heat conductivity.~\cite{LowTProp, LW_RMP, LW_BP} The plateau is
characterized by significantly enhanced phonon scattering, relative to
the two-level regime but not quite to the point of complete phonon
localization.  The term ``Boson'' refers to the linear temperature
dependence of the peak's strength.  Now, under certain circumstances,
which can be argued to be realised in glassy
semiconductors,~\cite{ZL_JCP, ZLMicro2} the boundaries of
reconfiguring regions also can host very special {\em electronic}
states of topological origin. All of these predictions are borne out
in experiment thus providing crucial support for the local description
in many experimental situations.

The article is organised as follows: the scenario of Yu and Leggett is
discussed in Section~\ref{infrange}. Section~\ref{local} covers the
picture based on the RFOT theory. The marginal stability scenario is
discussed in Section~\ref{marginal}.  The final Section~\ref{summary}
provides a summary of the main points and touches upon possible
experimental ways to differentiate between the various microscopic
scenarios that have been put forward.

\section{Yu-Leggett scenario}
\label{infrange}

In this Section, I review the main points of the Yu-Leggett
scenario. To set the stage, we note that the Green's function of a
massless field $\phi$ governed by the energy function $K (\nabla
\phi)^2/2$, in $d$ spatial dimensions, is given, up to a numerical
factor, by the expression $1/K r^{d-2}$, see for instance Section
5.7.5. of Ref.~\onlinecite{Goldenfeld}. Thus two elementary degrees of
freedom described indvidually by Eq.~(\ref{HTLS}) and located,
respectively, at $\br_i$ and $\br_j$, will be effectively coupled via
a ``dipole-dipole'' interaction
\begin{equation} J_{ij} \sim ({\bm g_i} \nabla)({\bm g_j} \nabla) /K
  r_{ij}^{d-2} \sim g^2/K r_{ij}^{d},
\end{equation}
up to a geometric factor reflecting the ``orientation'' of the
couplings ${\bm g}$ and the space dimensionality $d$:
\begin{equation} \label{Jijr} J_{ij} \: r_{ij}^d \propto \Gamma(d/2-1)
  \pi^{-d/2} d(d-2),
\end{equation}
see Eq.~(5.110) of
Ref.~\onlinecite{Goldenfeld}. Note the ${\bm g}$'s are vectorial for
the scalar elasticity $K (\nabla \phi)^2/2$ but must be rank-two
tensors in a proper treatment.~\cite{BL_6Spin, BLelast} For the sake
of argument we adopt the following functional form in three spatial
dimensions,
\begin{equation} \label{Jij0} J_{ij} \sim \frac{J}{r_{ij}^3},
\end{equation}
where 
\begin{equation} \label{Jij} J \equiv \frac{g^2}{K}.
\end{equation}

Suppose a macroscopic quantity of the elementary low-energy anharmonic
local degrees of freedom are embedded in a macroscopically stable
lattice. To avoid excessive repetition, we shall interchangeably call
these residual degrees of freedom ``elementary,'' ``bare,'' or
``primitive.'' Two such primitive degrees of freedom $i$ and $j$ will
form a resonance so long as their on-site energy difference is less
than $J/r_{ij}^3$. The latter coupling strength will also determine
the level spacing in such a resonance. At equilibrium below
temperature $T$, all resonant pairs separated by a distance shorter
than distance $r_T$ determined by
\begin{equation} \label{rT} k_B T \sim J/r_T^3,
\end{equation}
will be found largely in their quantum ground state, or
thermodynamically ``frozen-out.'' The {\em kinetics} of this
``freezing-out'' are not explicitly considered in this analysis but
could substantially affect measured quantities since measurements are
performed on finite timescales.  Thus below a temperature $T$, the
equilibrium concentration of thermally active resonant pairs is
limited from above by the quantity $1/r^3_T$ implying an upper bound
on the specific heat
\begin{equation}
  c(T) \lesssim  T/J
\end{equation} 
This can be restated as a bound on the density of states:
\begin{equation} \label{nE1} n(E) \lesssim \frac{1}{J},
\end{equation}
c.f. Ref.~\onlinecite{YuLeggett}. As just mentioned, the inequality in
the above equations reflects the uncertainty as to the spatial
arrangement of the resonant pairs, a notion we will return to
shortly. Perhaps more importantly, the inequality reminds us that the
argument is not entirely constructive: While it predicts the effective
density of strongly interacting degrees of freedom, it is agnostic as
to whether or why such degrees of freedom should be present in the
first place.

An improved estimate for the effective density of states of the
compound resonances was obtained by Yu and Leggett~\cite{YuLeggett}
who noted that the energy $E$ of a resonance will be statically
broadened as a result of freezing-out of those resonances exceeding
$E$ in energy by the amount
\begin{align} \label{DeltaE} \Delta E &\simeq \left| \Sigma_j J_{ij}
  \right| \simeq \rho_b \left| \int_{r_\smin}^{r_\smax} dr^3 (J/r^3)
  \right| \\
  & \simeq J \rho_b \left| \ln(r_\smax/r_\smin) \right| \\ & \simeq J
  \rho_b \ln(E_\smax/E) ,
\end{align}
where $\rho_b$ is the number density of the individual degrees of
freedom and the energy of a compound resonance is determined by $E
\simeq J/r^3$ as before. The ultraviolet cutoff energy $E_\smax$
should correspond to the closest spacing between individual elementary
degrees of freedom. The resulting density of states $\rho_b /\Delta E$
is, then:
\begin{equation} \label{nE2} n(E) \lesssim \frac{1}{J } \:
  \ln^{-1}(E_\smax/E).
\end{equation}
The above ideas were implemented systematically by Burin and
Kagan,~\cite{BurinKagan} see also Refs.~\onlinecite{Coppersmith,
  Carruzzo}. Note also that the logarithmic correction could
potentially explain rather effortlessly---even if not fully
quantitatively---why the heat capacity and conductivity should deviate
from the simple linear and quadratic dependences, respectively.

Thus Eq.~(\ref{nE1})---or its more accurate counterpart
(\ref{nE2})---provides an upper bound for the amount of instability in
any solid, a surprisingly simple yet apparently very general result.
What are physical implications of this general result?  Substituting
\begin{equation} \bar{P} = \frac{1}{J}
\end{equation}
in the ratio in the first equality in Eq.~(\ref{Uratio}) and using
Eqs.~(\ref{Jij}) and (\ref{nE1}), one obtains a lower bound on the
$(l_\smfp/\lambda)$ ratio for phonons:
\begin{equation} \label{llratio} l_\smfp/\lambda \gtrsim 1.
\end{equation}
This condition is exactly the venerable Ioffe-Riegel
criterion,~\cite{Mott1993} which prescribes that in order to define
semi-classically a quasi-particle, its mean free path must be greater
than its wavelength.  We note that a regime $l_\smfp/\lambda \simeq 1$
is in fact observed across all glassy substances in a broad
temperature range.~\cite{FreemanAnderson} (This range may not seem
broad on log-log plots~\cite{FreemanAnderson} but in fact, covers the
major portion of the temperature range that is ordinarily studied
below the glass transition.)

The upper bound on the density of states of the compound resonances in
Eqs.~(\ref{nE1}) and (\ref{nE2}) evidently becomes much less useful at
very low, sub-Kelvin temperatures. In this regime, this upper bound
exceeds by a factor of $10^2$ or so the apparent density of states
actually observed in experiment, c.f. Eq.~(\ref{Uratio}).  Recently
Vural and Leggett~\cite{VURAL20113528, doi:10.1021/jp402222g} have
presented calculations suggesting this issue may not be a
``deal-breaker'' but instead may only be a quantitative one in that it
seems to be satisfactorily resolved, if one explicitly specified that
the individual degrees of freedom be multi-level systems. Specific
molecular realisations of such multi-state resonances have been
proposed by Schechter and coworkers,~\cite{PhysRevLett.107.105504,
  PhysRevB.88.174202, 0953-8984-26-32-325401} in a variety of systems
such as disordered films of interest to quantum
computing.~\cite{Lisenfeld2017} One way to see how the
``multi-levelness'' affects the physics is to note that the couplings
$J_{ij}$ between the primitive degrees of freedom will generally scale
with the square of the number of levels while scattering of individual
phonons scales only linearly with that number.

In the Yu-Leggett scenario, the final density of states from
Eqs.~(\ref{nE1}) or (\ref{nE2}) does not depend on the detailed nature
of the individual elementary degrees of freedom or even on their
concentration $\rho_b$. The latter concentration only determines the
threshold ultraviolet energy below which the universality sets in: The
larger the concentration $\rho_b$, the higher the energy threshold.
Yet there is a potential issue posed by the type of density of states
in Eqs.~(\ref{nE1}) and (\ref{nE2}): It does not explicitly depend on
the material's particle density! To appreciate why this is
problematic, imagine the atoms were twice larger in size while the
coupling constant $J$ from Eqs.~(\ref{Jij}) did not change. Clearly
one now expects a density of states that is eight times lower, in
contradistinction with Eqs.~(\ref{nE1}) and (\ref{nE2}). Although the
elastic constants and other material constants are likely correlated
with the atomic size---which makes the preceding argument somewhat
delicate---the correlation is certainly not strict: A brief inspection
of a table of elastic moduli, for instance, reveals the moduli vary
within two or more orders of magnitude for conventional solids even
though the atomic size varies only within fifty percent or so. To
summarize, the correct scaling of the density of states in the
Yu-Leggett picture with the material density would be recoverd only if
there is a strict correlation between the coupling constant $g$ and
the elastic constant $K$ that involves the atomic size $r_\text{a}$:
\begin{equation} \label{cond} g^2 = C K r_\text{a}^3,
\end{equation}
where $C$ is a constant of dimensions energy that is not expressly
determined by the atom size.

Another potential complication has to do with the spatial arrangement
of the primitive degrees of freedom. Indeed, suppose for the sake of
argument the latter degrees were to comprise a {\em periodic} lattice
characterised by a perfectly spatially-uniform nearest neighbour
distance. If taken at its face value, the argument leading to
Eqs.~(\ref{nE1}) and (\ref{nE2}) suggests one would run out of
resonance-capable pairs above a certain, relatively short length and,
hence, below a certain energy. For instance, spin systems defined on a
periodic lattice have a gapped spectrum, or, if spin waves are
present, they exhibit a $T^{3/2}$ heat capacity at low temperature. In
contrast, the spectrum in Eq.~(\ref{nE2}) is significantly richer at
low energies. Of course, the couplings {\em are} expected to be random
in glasses, but how strongly random? In the extreme case of couplings
being distributed randomly around {\em zero}, we expect that the
effective molecular field acting on an individual degree of freedom is
no longer determined by the $J_{ij}$ themselves, as in
Eq.~(\ref{DeltaE}), but instead by their mean square
values.~\cite{TAP, PWA_LesHouches} The integral over $J^2 \propto
1/r^6$ is perfectly infra-red convergent thus obviating the argument.
 
One way to include effects of the spatial arrangement of the primitive
degrees of freedom would be to reformulate the argument leading to
Eqs.~(\ref{nE1}) and (\ref{nE2}) in the {\em reciprocal} space. We
have seen that to avoid a gap in the density of states of the
resonances, it is imperative that the bare degrees of freedom do not
comprise a strict lattice. Instead their concentration must exhibit
deviations from the strictly periodic or uniform spatial arrangement
down to the lowest wavevector $q$. The magnitude of such deviations on
the length scale $r \sim 1/q$ is given by the structure factor of the
system: $S(q) \equiv (1/N) \sum_{ij} e^{iq(r_i - r_j)}$. In view of $E
\sim J q_T^3$, one obtains
\begin{equation} \label{nE3} n(E) \lesssim S[(E/J)^{1/3}] \:
  \frac{1}{J},
\end{equation}
where we have omitted the logarithmic correction from Eq.~(\ref{nE2})
partially because the rate of change of the structure factor at
relevant energies well exceeds that of the logarithm and partially for
simplicity. We note that $S(q \to \infty) \to 1$ thus recovering the
high energy asymptotics from Eqs.~(\ref{nE1}) and (\ref{llratio}).

The low energy asymptotics are more interesting. In the $q \to 0$
limit, the structure factor is determined by the typical magnitude of
density fluctuations (p.~313 of Ref.~\onlinecite{L_AP}) and, hence, by
the compressibility: $S(q \to 0) \to \rho k_B T/K$ where $\rho$ is the
particle density. Since the structure of the glass is largely arrested
below the glass transition temperature $T_g$, except for some ageing
and subtle changes related to a decrease in the vibrational amplitude,
we thus obtain:
\begin{equation} \label{nE4} \bar{P} \lesssim \frac{\rho k_B T_g}{K}
  \: \frac{1}{J}.
\end{equation}
Note the above expression now does contain the material's density,
but, seemingly, at the expense of universality. Recall, however, that
by the equipartition theorem, $K \la (\nabla \phi)^2 \ra = k_B T$.  At
the same time, the relative displacement $\la |\nabla \phi| \ra$
corresponds with the typical vibrational displacement $d_v$ relative
to the particle spacing $a$: $\la |\nabla \phi| \ra =
d_v/a$. Therefore we have
\begin{equation} \label{equip} \left( \frac{d_v}{a} \right)^2 \simeq
  \frac{\rho k_B T_g}{K},
\end{equation}
leading to 
\begin{equation} \label{nE5} \bar{P} \lesssim
  \left(\frac{a}{d_v}\right)^2 \frac{1}{J}.
\end{equation}
The ratio $d_v/a$ is bounded from above by and numerically close to
the ratio $d_L/a$, where the quantity $d_L$ is the so called Lindemann
displacement, i.e., the vibrational displacement at conditions where
mechanical rigidity sets in. Numerically, $d_L/a \simeq 0.1$ almost
universally in the physical three dimensions, which is the venerable
Lindemann criterion of melting.~\cite{L_Lindemann, RL_LJ} Thus, at
sufficiently low temperatures,
\begin{equation} \label{UratioYL} \frac{l_\smfp}{\lambda} \simeq
  \left(\frac{g^2}{J K} \right)^{-1} \left(\frac{a}{d_L}\right)^2
  \gtrsim 10^2,
\end{equation}
nearly universally, in view of Eq.~(\ref{Jij}). The formula above
requires that the phonon scattering be resonant. This affords one an
internal consistency check within the Yu-Leggett scenario, since the
condition for resonant scattering, $r_T < \lambda$, implies the
``weakly-dampened'' regime in Eq.~(\ref{UratioYL}) can be observed at
temperatures such that:
\begin{equation} \label{JT} k_B T \lesssim (k_B T_D a)^{3/2}/J^{1/2},
\end{equation}
where $T_D$ is the Debye temperature and we used Eq.~(\ref{rT}). Since
$r_T \ge a$, one obtains a very reasonable upper bound
\begin{equation} \label{Jinf} J < k_B T_D a^3 \Rightarrow E_\smax
  \simeq k_B T_D,
\end{equation}
up to a numerical factor of order one.  Leggett~\cite{Leggett} pointed
out that Eq.~(\ref{JT}) defines a temperature below which the
interacting defects can be thought of as Ising-like spins, because
resonant exchange of phonons allows for flip-flop processes, while
above that temperature, the model is more Heisenberg-like.

Thus we observe that a microscopic picture of a macroscopically stable
amorphous solid, in which strongly interacting local anharmonic
degrees of freedom form resonant, spatially extended pairs appears to
be internally consistent. The density of states of the resonances
drops by a factor of $10^2$ or so once their spatial extent
significantly exceeds the molecular length. In both high and low
energy limits, the $l_\smfp/\lambda$ is nearly universal and
numerically close, respectively, to $10^0$ and $10^2$. The appearance
of the ratio $(a/d_L)^2$ in the low-energy part of the density of
states (\ref{UratioYL}) is significant: In modern theories of
liquid-to-solid transitions,~\cite{L_AP} this ratio plays the crucial
role of the order parameter for translational symmetry breaking. Its
magnitude is of order $10^2$ in solids but changes discontinuously to
zero in the liquid. It is reassuring that Eq.~(\ref{nE5})
automatically implies no frozen-in degrees of freedom could be present
in a uniform liquid.  Finally note the above argument can be
reformulated in all dimensions $d \ge 3$. The $(a/d_L)$ ratio is
expected to increase linearly with the space
dimensionality:~\cite{MCT}
\begin{equation} \label{adL} \frac{a}{d_L} \propto d,
\end{equation} 
while the coupling constant increases combinatorially rapidly with
$d$, as remarked just below Eq.~(\ref{Jijr}). Consequently, the
numerical value of the lower bound (\ref{UratioYL}) quickly decreases
to the overdamped value of $1$ even for dimensions only modestly
exceeding the physical value $3$. This limits the usefulness of this
lower bound on the $l_\smfp/\lambda$ ratio for putative
high-dimensional solids. In any event, the Yoffe-Riegel bound in
Eq.~(\ref{llratio}) reminds one of the original Einstein's view on
solids as collections of uncorrelated oscillators while heat transfer
is carried out by short wave packets,~\cite{Einstein} a view we now
recognise as meanfield. Note that according to Eq.~(\ref{DeltaE}), the
interaction volume is determined the system's volume itself. In this
sense, the interactions are infinite-range.

Despite its generality, the infinite-range scenario is agnostic as to
whether the bare, local degrees of freedom should be present in the
first place. While this lack of being constructive may seem an
attractive feature of the argument, it does call for some
vigilance. Indeed, there are plenty of amorphous materials that either
do not exhibit detectable TLS-like degrees of freedom, such as freshly
deposited silicon films or where two-level density seems to depend on
preparation.~\cite{Pohl_review, QeenHellman2015, McKennaAmber}
(Nevertheless, such films progressively develop an excess of local
resonant degrees of freedom when annealed, in the absence of
crystallisation.) By construction, the density of states for the bare
degrees of freedom should be higher than that prescribed by
Eq.~(\ref{nE3}). Conversely, a set of {\em local} degrees of freedom
characterised by the DOS from Eq.~(\ref{nE3}) will not form
resonances.~\cite{Silbey} Next we turn to a scenario that does
describe on a constructive basis just how local degrees of freedom
could emerge in an aperiodic solid made by a quench of an equilibrated
liquid.


\section{The mosaic scenario and its quantization}
\label{local}

We first review select notions of the non-meanfield scenario of how
mechanical rigidity emerges in structural glasses that is provided by
the random first order transition (RFOT) theory. An up to date and
very extensive, pedagogical account of the RFOT theory~\cite{L_AP} as
well as an older, less technical review~\cite{LW_ARPC} are
available. A recent article~\cite{LWjamming} expands those notions
established for molecular systems to a greater variety of preparation
protocols so as to include amorphous collections of particles such as
jammed colloidal particles or granular assemblies.

The RFOT theory generalizes some established, quantitative notions of
the liquid-to-{\em periodic}-crystal transition. The latter transition
represents the extreme case where the translational degrees of freedom
freeze out all at the same time in a first order transition, thus
leaving in essentially no degrees freedom other than lattice
vibrations. Important and interesting exceptions to this notion
include solids that are largely periodic but that still house strongly
anharmonic structural excitations such as phasons,~\cite{Bak1982}
orientational dynamics in relaxor
ferroelectrics,~\cite{doi:10.1080/00018732.2011.555385} or proton
transport in solid ice,~\cite{doi:10.1021/ja01315a102} to name a few.

What makes liquid-to-solid transitions difficult for analytical
treatment is that they are intrinsically, strongly discontinuous thus
rendering power-law expansions for small values of the order parameter
mathematically dangerous. The necessity for the discontinuity itself
was elegantly argued for by Landau already in 1937,~\cite{LandauPT1}
who showed that a liquid-to-solid transition could be continuous at
most in one isolated phase point. Brazovsky showed decades later that
if such a critical point existed, it would be pushed down to absolute
zero by fluctuations.~\cite{Brazovskii1975, PhysRevLett.41.702} Also
in 1937, Bernal~\cite{Bernal1937} made a much less formal yet equally
insightful notion that the entropy of a liquid can be presented as a
sum of translational and vibrational entropy---a view justified for
covalently bonded liquids~\cite{L_AP}---implying crystallisation
always exhibits latent heat.

Early attempts at developing quantitative theories of {\em periodic}
crystallisation by Kirkwood and others~\cite{KirkwoodMonroe} followed
a Landau-like approach and employed, as order parameters, expansion
coefficients in front of the Fourier components of the solid's density
profile. A non-vanishing coefficient in front of a non-zero $q$
Fourier component means translational invariance is broken.  These
attempts achieved much conceptual progress~\cite{RyzhovTareeva} but
being limited to only a few Fourier components for technical reasons,
fell short of producing a quantitative theory.  A technical
breakthrough in the field was achieved when a different density ansatz
for the solid was adopted, viz., a sum of Gaussians:
\begin{equation} \label{rhoalpha} \rho(\br) = (\alpha/\pi)^{3/2}
  \sum_i e^{-\alpha(\br - \br_i)^2},
\end{equation}
where the summation is over the vibrationally averaged positions of
individual particles. Note this ansatz would be exact for a strictly
harmonic solid comprised of equivalent particles.~\cite{LLstat,
  RL_Tcr}

A class of non-perturbative density functionals were subsequently
found that happen to sum an infinite subset of diagrams~\cite{MWDA}
and that quantitatively reproduce the phase behaviours of hard spheres
and soft particles alike.~\cite{PhysRevLett.56.2775} In these
approaches, the optimal value of the order parameter $\alpha$ is found
variationally, by optimising the free energy $F(\alpha, \bar{\rho})$
with respect to both $\alpha$ and the average density $\bar{\rho}$.
Felicitously, the uniform liquid is {\em also} described by the
density ansatz (\ref{rhoalpha}) if one sets $\alpha = 0$, thus
allowing one to build an effective free energy function that applies
to both phases and, significantly, to the inter-phase transition
state.  The order parameter $\alpha$ changes its value from zero to
$10^2/a^2$ following crystallisation, a discontinuous transition
indeed. A non-zero value of $\alpha$ implies that each atom is now
subject to an effective confining potential in the form of an Einstein
oscillator. The analysis can be extended, in periodic crystals, to
small deformations around vibrational ground states and thus can be
used to test for mechanical stability in a broad range of
wavelengths. Note that the value of $\alpha$ near the saddle point of
the free energy $F(\alpha, \bar{\rho})$ corresponds with the inverse
square of the vibrational displacement near the stability limit:
$\alpha \simeq 1/d_L^2$, up to a numerical constant of order one,
c.f. Eq.~(\ref{UratioYL}).

In an equally quantitative fashion, it has been shown that a generic
{\em aperiodic} lattice can be metastable with respect to the uniform
liquid, whether it is made of hard or soft particles.~\cite{dens_F1,
  BausColot, RL_LJ} When aperiodic metastable minima of the free
energy functional emerge, the free energy excess of a specific {\em
  individual} minimum over the uniform liquid ensemble is typically
around $k_B T$ per particle. At the same time, the number of distinct
individual aperiodic minima scales exponentially with the system size,
$e^{s_c N/k_B}$, where the quantity $s_c$, called the configurational
entropy, is numerically close to $k_B$. Indeed, the excess entropy of
a liquid relative to the corresponding crystal is empirically known to
be about $1.5 k_B$ per atom or larger near melting.~\cite{CRC} That
the free energy deficit of an individual free energy minimum is
compensated by the entropic stabilization due to the multiplicity of
the minima implies that the totality of the distinct aperiodic
structures could be just as stable as the uniform liquid. {\em
  Computation} of the configurational entropy for actual liquids is
difficult but has been accomplished relatively recently for model
liquids within the replica-symmetry breaking (RSB)
framework.~\cite{mezard:1076} The latter framework has an important
methodological dividend: The totality of all metastable structures
described by the ansatz (\ref{rhoalpha}) automatically turns out to
have the same free energy and {\em total} entropy as the uniform
liquid. (This is similar to what is seen in a variety of related
spin-models.~\cite{MCT1}) And so, just the sole fact of emergence of
metastable aperiodic minima automatically implies that the minima
correspond with equilibrium configurations. To contrast aperiodic
crystal formation from the ordinary liquid-to-crystal transition, the
emergence of the degenerate aperiodic solids has come to be called a
{\em random} first order transition (RFOT). As just mentioned, the
entropy of the liquid is a smooth function of the temperature; thus
unlike the transition to a periodic crystal, {\em the RFOT has no
  latent heat}.

\begin{figure}[t] \centering
  \includegraphics[width= \figurewidth]{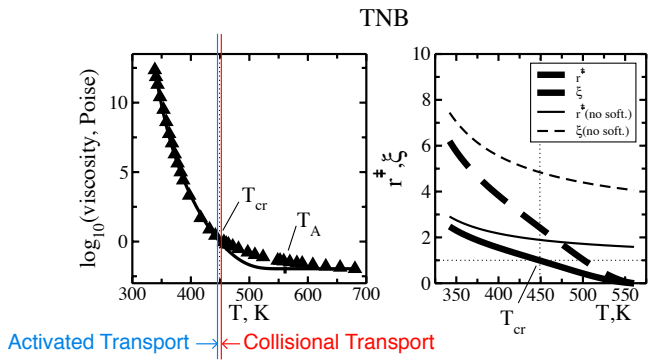}
  \caption{\label{crossover} The left panel displays
    experimentally-determined temperature dependence of the viscosity
    of an organic glass-former TNB (symbols). Also shown is the
    theoretical prediction~\cite{LW_soft} for the relaxation time for
    activated transport, Eq.~(\ref{Fdagger}), also accounting for the
    effects of fluctuations of the order parameter $\alpha$ from
    Eq.~(\ref{rhoalpha})~\cite{LW_soft} but largely ignoring
    mode-coupling effects. The crossover corresponds to the
    temperature at which experimental data diverge from the
    theoretical prediction. The right panel compares RFOT-based
    predictions of the critical radius $r^\ddagger$ and cooperativity
    length $\xi$ with and without including the ``barrier-softening''
    effects~\cite{LW_soft} due to fluctuations of $\alpha$. Note the
    crossover happens to coincide with the temperature at which the
    critical radius for structural reconfigurations is numerically
    close to the molecular length scale, $r^\ddagger \approx a$.}
\end{figure}

The emergence of transient rigidity predicted by the RFOT theory---in
the form of long-lived local structures---immediately explains the
established experimental fact that below a certain temperature (or
above a certain density), transient cages form around each molecule
already in an {\em equilibrated} liquid as directly seen, for
instance, in neutron scattering~\cite{MezeiRussina} where one finds an
extended plateau in the intermediate scattering function; the
Debye-Waller factor of a transiently trapped particle can be directly
measured and is numerically close to what it would be in a perfectly
stable solid.  The lifetime $\tau$ of the cage can be arbitrarily
long---only subject to one's patience---and ultimately determines the
viscosity $\eta$ of the liquid according to:~\cite{Frenkel, Lionic,
  LLelast}
\begin{equation} \label{Maxwell} \eta \simeq \mu \tau,
\end{equation}
originally due to Maxwell; $\mu$ is the high-frequency shear
modulus. In addition to the aforementioned plateau observed in neutron
scattering experiments, the formation of transient cages leads to
other consequences that have been observed, such as the violation of
the Stokes-Einstein relation or decoupling of translational and
rotational diffusion, and decoupling of dielectric and mechanical
relaxation, see Refs.~\onlinecite{XWhydro, Lionic, RL_Tcr} and
references therein.  Microscopic arguments indicate this transient
ergodicity breaking is a soft crossover occurring near viscosities of
order $10^1-10^2$~Ps,~\cite{LW_soft, SSW} see Fig.~\ref{crossover}, a
result consistent with experiment. At this viscosities, the activation
energy of transport begins to grow rapidly. The crossover, whose
ballpark temperature we denote with $T_\scr$, becomes a sharp
transition in the meanfield limit. The temperature of the latter
meanfield transition is often denoted with $T_A$~\cite{MCT, MCT1} or
the dynamical temperature $T_d$.

Thus we arrive at a seeming conflict: On the one hand, the free energy
of the liquid below the crossover is locally minimised by a spatially
non-uniform density profile (\ref{rhoalpha}) with $\alpha \simeq
10^2$, where the site locations $\{ \br_i \}$ now correspond with an
aperiodic lattice. On the other hand, the liquid flows, thus
eventually restoring the translational symmetry and implying that
$\alpha = 0$ on long times. How can one have it both ways at the same
time with regard to the value of $\alpha$?

The situation is indeed somewhat subtle, but actually does not present
a contradiction. To appreciate this, imagine first an individual,
small, and isolated ferromagnetic domain sufficiently below the Curie
temperature of the bulk material so that the domain maintains its
magnetisation long enough for the observer to notice and measure it,
but not too long so that the polarization of the domain will typically
flip before the observer runs out of patience. (Magnetic recordings
deteriorate exactly because of such flips.) Thus, on the one hand, the
free energy is minimised in one of the two polarised states: The ratio
of the duration of the flip (i.e. the instanton time) to the typical
wait time inside a minimum can be made arbitrarily small. On the other
hand, the {\em average} magnetisation is still zero and the symmetry
is maintained but only on very long times. Now imagine a {\em
  macroscopic} magnet below the Curie point, but not too much lower,
and assume for convenience that the magnet was cooled so that the
average magnetisation is small, much much lower than what it would be
in a fully polarised sample. Clearly, the sample will represent a
``mosaic'' of regions polarised up or down. The so polarised domains
are separated by domain walls. The number, extent, and shape of the
walls will depend on the mismatch penalty between the distinct
polarisations, translational and vibrational entropy of the walls,
etc.  The mosaic is not static because of the non-vanishing
translational and vibrational entropy of the domain walls. Thus for
long times, the magnetisation of the sample as a whole and any of its
regions will be zero {\em on average}. At the same time, the free
energy of the sample is minimised by polarised solutions, not the
paramagnetic solution!

Likewise, a liquid can be thought of as a mosaic made of distinct free
energy minima when $\alpha > 0$.~\cite{KTW, XW} However in contrast
with the ferromagnet, which has only two distinct free energy minima
to speak of, the liquid has exponentially many such minima. In this
case, the lengthscale of the mosaic is determined much less by the
entropy of the domain walls than by the multiplicity itself of the
minima. Calculations~\cite{EastwoodW} suggest the the multiplicity of
the minima dominates the pertinent physics for values of the
configuration entropy that exceed already the very small value of
$10^{-4} k_B$ per particle.  Sufficiently below the temperature when
the minima begin to form, local interconversions between the minima
can be considered as rare events so that their rates can be accurately
estimated using the venerable transition state theory.~\cite{FW,
  Hanggi_RMP} Specifically, the free energy cost of locally replacing
the existing configuration by another, equivalent aperiodic
configuration can be presented as a sum of two
contributions:~\cite{KTW}
\begin{equation} \label{FN1} F(N) = - T s_c N + \gamma N^{1/2},
\end{equation}
where $- T s_c$ is the bulk driving force and the odd-looking term
$\gamma N^{1/2}$ approximates the mismatch penalty between two
distinct aperiodic free energy minima. The scaling of this mismatch
penalty is slower, in dimensions three or higher, than the
$N^{(d-1)/d}$ scaling expected for interfaces between two distinct
thermodynamic phases. This reduction comes about because a strained
region separating two equivalent aperiodic minima can typically lower
its energy by as much as $\sim N^{1/2}$ by locally replacing any
region by yet another equivalent random structure.~\cite{KTW, XW,
  LRactivated, CL_LG} Appropriately, the resulting stabilisation of
the interface scales with the magnitude of Gibbs free energy
fluctuations of an individual minimum~\cite{KTW, LRactivated} and is
thus approximately determined by the bulk modulus of the
material:~\cite{LRactivated, L_AP}
\begin{equation} \label{gamma2} \gamma \approx (K k_B T/ \rho)^{1/2}.
\end{equation}
This and other available detailed estimates for the coefficient
$\gamma$ yield values for relaxation barriers in glassy liquids that
agree well with experiment,~\cite{XW, RL_sigma0, RWLbarrier,
  LRactivated} without using any adjustable parameters.  The kinetic
barrier corresponding to the free energy profile (\ref{FN1}),
\begin{equation} \label{Fdagger}
  F^\ddagger = \gamma^2/4 k_B T s_c, 
\end{equation}
is finite unless the configurational entropy vanishes. This barrier
directly determines the rate at which the liquid will flow when
sheared, by virtue of $\tau^{-1} = \tau_0^{-1} e^{-F^\ddagger/k_B T}$
and Eq.~(\ref{Maxwell}).  The quantity $\tau_0$ is a vibration
timescale numerically close to a picosecond.~\cite{LW_Wiley} The
critical size corresponding to the profile (\ref{FN1}),
\begin{equation} \label{Ncrit}
N^\ddagger = (\gamma/2 T s_c)^2, 
\end{equation}
is finite and obligingly becomes greater than the molecular size 1
below the crossover temperature $T_\scr$,~\cite{LW_soft, LWjamming}
Fig.~\ref{crossover}(b), consistent with the premise that each
individual particle is confined to a cage. One may further ask: What
is the size $N^*$ of a region that will have reconfigured once the
critical size $N^\ddagger$ is reached?  The reconfiguration size $N^*$
is an important quantity because it corresponds with the lengthscale
of the mosaic and determines the cooperativity length for activated
transport.  To answer the above question, we first recall that in the
expression (\ref{FN1}) for the free energy cost of reconfiguring $N$
particles, the energy reference is set at zero, $F(N=0)=0$. An equally
likely configuration of the same sample but with a region of size
$N^*$ flipped is thus determined~\cite{XW, LW, LW_aging, LRactivated}
by a non-vanishing solution $N^*$ of the equation $F(N)|_{N=N^*}=0$.
The quantity $N^*$ therefore gives a lower bound for the size of a
region that can be occupied by a single aperiodic minimum.~\cite{LW,
  LW_aging, BouchaudBiroli} In fact, it is also the {\em typical} size
occupied by a single solution since beyond $N^*$, there is a
non-vanishing probability that yet another solution will be present in
the region: $-F(N)|_{N>N^*}/k_B T > 0$. Eqs.~(\ref{FN1}) and
(\ref{gamma2}) yield straightforwardly
\begin{equation} \label{N*}
  N^* = \frac{\gamma}{T s_c}^2 = \frac{K}{\rho k_B T (s_c/k_B)^2}.
\end{equation}
To gain an additional perspective on this result, one may recall the
expression for the particle-number fluctuation in a {\em uniform}
liquid:~\cite{LLstat} $\delta N = (N \rho k_B T/K)^{1/2}$. This means
that the lower bound on the size $N^{(1)}$ of a region which will
routinely accommodate for adding/removing one particle particle,
$\delta N = \pm 1$, is given by $N^{(1)} = K/\rho k_B T$, c.f. the
second equality in Eq.~(\ref{N*}). Yet this type of ``free
volume''~\cite{PhysRevB.20.1077} mechanism for mass transport is
irrelevant for systems with broken translational symmetry such that
the density profile is given by Eq.~(\ref{rhoalpha}), $\alpha \simeq
10^2$. Under the latter circumstances, the liquid can no longer flow
and so creation of free volume would require creating a {\em
  vacancy}. The latter would cost many multiples of the thermal
energy,~\cite{LWisotope} $\sim K a^3 \simeq \alpha k_B T \simeq 10^2
k_B T$, c.f. Eq.~(\ref{equip}), and thus occurs very rarely. The
cooperativity length for creating a vacancy would be strictly
infinite, by Eq.~(\ref{N*}), since disallowing motions other than
vibrations implies $s_c = 0$.

Now, in three dimensions the physical length corresponding to the
cooperativity size $N^*$ is given by:
\begin{equation} \label{xi} \xi \equiv a (N^*)^{1/3} = a (\gamma/T
  s_c)^{2/3}.
\end{equation}
The cooperativity size $N^*$ depends algebraically on the barrier and,
hence, only logarithmically on the relaxation time. Indeed,
Eqs.~(\ref{Fdagger})-(\ref{xi}) yield:
\begin{equation} \label{Ntau} N^* \equiv \left(\frac{\xi}{a}\right)^3
  = \left(\frac{4F^\ddagger}{\gamma}\right)^2 \simeq 16 \frac{k_B T
    \rho}{K} \ln^2(\tau/\tau_0).
\end{equation}

Because the timescale for conventional quenches varies between seconds
and hours while $\tau_0 \simeq 10^{-12}$~sec, we conclude that the
cooperativity size is apparently nearly universal for routinely made
glasses.  Numerically, this size is about 200,~\cite{XW, LW,
  RL_sigma0, RWLbarrier, LRactivated} implying
\begin{equation} \label{xia6}
  (\xi/a) \simeq 6,  \text{ at } T = T_g. 
\end{equation}
For speedier quenches, the cooperativity size $N^*$ can be made
significantly smaller than 200, but probably no smaller than a few
tens of particles, see Ref.~\onlinecite{LL1} and also the discussion
of cooling protocols in Section~\ref{marginal}. Since the molecular
size $a$ corresponds with the size of a rigid molecular unit, or
``bead,''~\cite{LW_soft} which typically measures several angstroms
across,~\cite{LW_soft, LL1} the cooperativity length $\xi$ reaches
only a few nanometers in magnitude at the glass transition.  This is
consistent with indirect observations using several types of
non-linear spectroscopy~\cite{Spiess, RusselIsraeloff, CiceroneEdiger}
as well as direct observation of cooperative reconfigurations on the
surface of metallic glasses.~\cite{GruebeleSurface} We note that the
cooperative reconfigurations amount to a dynamic, not static
heterogeneity. To detect dynamic heterogeneity, one must measure a
four-point correlation function, which is not accessible to linear
spectroscopy. Conversely, no easily describable static structural
patterns would be obvious on the length scale
$\xi$,~\cite{0295-5075-98-3-36005} see however below for possible
electronic signatures.

To estimate the absolute values of $a$ and $\xi$---as opposed to the
dimensionless ratio $\xi/a$---one must be able to identify the
effective particle of the theory, i.e. the bead. Rigid molecular units
are straightforwardly identified in molecular substances; the
corresponding size matches well its value as determined by calibrating
the entropy of fusion of the substance by that of a Lennard-Jones like
substance.~\cite{LW_soft} Such chemical determination however becomes
ambiguous in covalently bonded liquids such as the chalcogenide
alloys. Very recently, Lukyanov and Lubchenko~\cite{LL1} (LL) have
designed a chemically-inspired algorithm to generate ensembles of
octahedrally coordinated amorphous structures that turn out to
reproduce consistently the first sharp diffraction peak (FSDP) in the
structure factor.  The FSDP is viewed by many as reflecting the so
called medium-range order in glasses and has been a subject of debate
for decades.~\cite{ISI:A1994NL31200006, ElliottNature1991, Mei2008,
  PhysRevB.92.134206} LL~\cite{LL1} have argued that the lengthscale
underlying the FSDP is, in fact, the bead size. Thus the basic notions
of the RFOT theory can be connected directly with actual substances
while allowing one to obtain estimates for absolute values of the
reconfiguration length and barrier.~\cite{RWLbarrier, LRactivated}
Generically,~\cite{L_AP}
\begin{equation} \label{xiconc} 1/\xi^3 \simeq 10^{20} \text{
    cm}^{-3},
\end{equation}
i.e. one domain per several hundred atoms.

Thus Eqs.~(\ref{FN1})-(\ref{Ncrit}) mathematically embody a key
physical notion regarding the emergence of mechanical stability in
aperiodic solids: The material is stable on a finite length scale
$\xi$, which meaningfully exceeds the molecular length, while larger
regions are only metastable, the escape barrier given by
Eq.~(\ref{Fdagger}). Thus the question of the mechanical stability of
the system is that of quantifying the latter escape barrier. According
to Eqs.~(\ref{Fdagger}) and (\ref{xi}), the reconfiguration barrier
and extent both increase with cooling because the configurational
entropy, as any entropy, must decrease with temperature. A very good
empirical fit is given by the functional form~\cite{RichertAngell}
\begin{align}  s_c &= \Delta c_p(T_g) T_g (1/T_K-1/T) \nonumber \\
  &\propto (T-T_K), \hspace{5mm} T \to T_K, \label{scT}
\end{align}
where $\Delta c_p(T_g)$ is the heat capacity jump at the glass
transition. The resulting temperature dependence of the relaxation
barrier and cooperativity length are exemplified in
Fig.~\ref{crossover}.

The stability of aperiodic structures made by quenching a liquid is
thus achieved not because such an aperiodic state is strictly unique,
in contrast with ordinary crystalline solids, but because the
alternative states are behind sufficiently high barriers and their
number is sufficiently small. How stable are glasses?  At the glass
transition achieved using a routine quench, $\tau/\tau_0 =
e^{F^\ddagger(T_g)/k_B T_g} \simeq 10^{15}$. At a temperature half as
large as $T_g$, the corresponding relaxation time would be
$\tau/\tau_0 = e^{2 F^\ddagger(T_g)/k_B T_g} \simeq 10^{30}$, a
monstrously large number. In reality, relaxation in frozen glasses
occurs toward states that are stabilized enthalpically relative to the
frozen-in states and is much much faster than that implied by the
naive estimate above.~\cite{LW_aging} Nevertheless, the relaxation
times for ageing in deeply-quenched glasses are still astronomically
large,~\cite{LW_aging} consistent with the everyday experience that
glasses are often mechanically harder than periodic crystals, even if
at the expense of being brittle.

The emergence of the degenerate aperiodic crystal characterised by the
density profile (\ref{rhoalpha}) is largely driven by steric
repulsion;~\cite{L_AP} lower temperature and/or increase in pressure
is simply a means to increase density. Elastic response, at non-zero
frequencies, then arises self-consistently: It can be shown
straightforwardly that $\alpha$ is equal to the shear modulus of an
individual aperiodic minimum, up to a numerical constant determined by
the Poisson ratio.~\cite{RL_Tcr} The large value of order parameter
$\alpha$ ($\alpha a^2 \simeq 10^2$) explains, in retrospect, why
perturbative treatments of liquid-to-solid transitions are
mathematically difficult. The corresponding, small vibrational
amplitude $1/\alpha^{1/2}$ thus can be regarded as the key emergent
quantity in the problem. Appropriately, this emergent (very small!)
length scale readily arises in meanfield and non-meanfield treatments
alike. Because each individual reconfiguration must occur just at the
mechanical stability edge for the moving atom---so as not to
compromise the stability of the lattice--- the length scale
$1/\alpha^{1/2}$ also has the clear physical meaning~\cite{XW} of the
displacement of an individual particle during a reconfiguration: $d_L
= \alpha^{-1/2}$. This notion also formally enters in RSB
treatments~\cite{mezard:1076, MP_Wiley} because two replicas are
considered distinct if their overlap, which is basically a
Debye-Waller factor of sorts, is less than a certain value.

\begin{figure}[t] \centering
  \includegraphics[width= .85 \figurewidth]{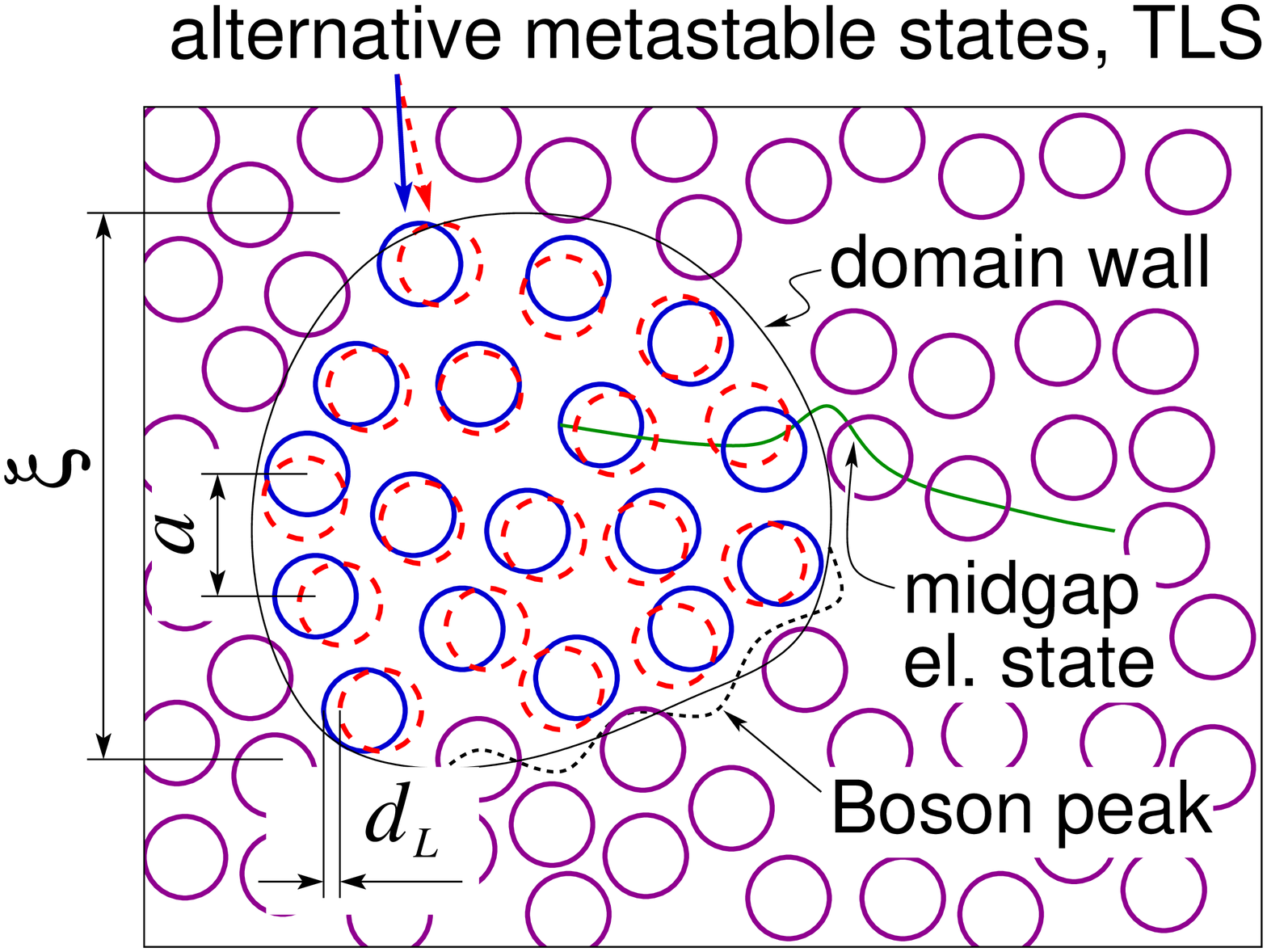}
  \caption{\label{domain} Partial graphical summary of some of the
    physics associated with the mosaic picture of an equilibrated
    glassy liquid below $T_\scr$. The quantities $a$, $d_L$, and $\xi$
    stand, respectively, for the lattice spacing, the particle
    displacement at the mechanical stability edge, and the
    cooperativity length for structural reconfigurations. The
    intrinsic uncertainty in the position of the boundary can be
    thought of as vibrations of the domain wall encompassing the
    reconfiguring region; these vibrations account for the Boson
    peak.~\cite{LW_BP} The tension of the wall has a contribution from
    special midgap electronic states that are present, if the material
    exhibits spatially-varying bond saturation.~\cite{ZL_JCP,
      ZLMicro2} }
\end{figure}

The lengthscale $\xi$, on the other hand, is intrinsically
non-meanfield; it arises only because of the {\em locality} of
interactions. It turns out the latter non-meafield effects undergird
the great majority of the dozens of quantitative predictions the RFOT
theory has made, see reviews in Refs.~\onlinecite{LW_ARPC, L_AP}. Of
particular significance in the present context are the {\em
  quantum-mechanical} consequences of the special physics associated
with the length $\xi$, which are often overlooked, perhaps because the
translational symmetry breaking $\alpha=0 \to \alpha>0$ that occurs as
a result of the random first order transition is entirely of a
classical origin for most substances known. (Some interesting new
physics could appear for helium~\cite{RevModPhys.84.759} or
electrons!)  Here we highlight several features that have been
reviewed previously in Refs.~\onlinecite{L_JPCL, L_AP} and, in
addition, some more recent results.

To quantise the structural excitations of a frozen solid that is a
mosaic of aperiodic free energy minima, we first enquire about the
classical density of states of such a solid just above a {\em kinetic}
glass transition at a temperature $T_g$. The word ``kinetic'' implies
that there are still lower energy states that would have been
available in principle to the system, if it had been cooled more
slowly. Nevertheless, the system became kinetically arrested in an
off-equilibrium configuration because the cooling rate was greater
than the inverse relaxation time of a subset of the translational
degrees of freedom. If, on the other hand, no lower-energy states were
available, then the glass transition would be truly thermodynamic. The
putative lowest energy aperiodic state is often called the ideal,
Kauzmann state.~\cite{Kauzmann} Now, the smallest reconfigurable
region has exactly one state available near $T_g$. (As far as the
corresponding reconfiguration is concerned, the surrounding matrix is
by construction regarded as purely elastic.)  Thus with regard to the
latter region, the temperature $T_g$ corresponds to a {\em
  thermodynamic} glass transition. A general form of the density of
states for such a system, at low energies, is~\cite{MPV,
  SpinGlassBeyond, Derrida} $\Omega(E) \propto \exp[E/k_B T_g +
O(E^2)]$, where the ground state is at $E = 0$ by construction. The
glass transition formally comes about because in a {\em finite}
temperature range $T \in [0, T_g]$, the integrand of the partition
function $\int dE \: \Omega(E) \: e^{-\beta E}$ is dominated by the
vicinity of $E=0$. That is, the system freezes into its ground state
already at the finite temperature $T_g$. This formally corresponds to
a density of states $\Omega(E) = (1/k_B T_g) \exp(E/k_B T_g)$, the
prefactor needed to reflect that exactly one ground state is found at
energy $E = 0$ or below: $\int_{-\infty}^0 dE \: \Omega(E) =
1$. Considering that at any given time the spatial concentration of
regions that can reconfigure is $1/\xi^3$, one obtains that the total
density of states due to the domains is
\begin{equation} \label{nEclass} n(E) \simeq \frac{1}{k_B T_g \xi^3}
  e^{E/k_B T_g}.
\end{equation}
Thus without explicit consideration of the kinetics of structural
excitations accessible at cryogenic temperatures, the above expression
yields for the density of states of low energy excitations: $\bar{P} =
1/k_B T_g \xi^3$, since $T_g$ of most glasses is much larger than
liquid helium temperatures. Substituting specific values for the glass
transition temperature and bead size $a$ yields $\bar{P} \simeq
10^{45\pm1}$~m$^{3}$J$^{-1}$. This value is consistent with
experiment.~\cite{BerretMeissner}

In addition, optimising the phonon part of the TLS $+$ phonon energy
function, $\sigma_z {\bm g} \nabla \phi + \int d^3\br K(\nabla
\phi)^2/2$, with respect to the displacement yields $\sigma_z {\bm
  g}/a^3 = - K \nabla \phi$. Multiplying this by $\nabla \phi$,
recalling that $\la K(\nabla \phi)^2/2 \ra = k_B T/2$ by
equipartition, and noting that $|\la \sigma_z \nabla \phi \ra| \simeq
| \nabla \phi|$, one obtains
\begin{equation} \label{g} g \simeq (K k_B T/\rho)^{1/2},
\end{equation}
c.f. Eqs.~(\ref{cond}) and (\ref{gamma2}). According to the discussion
leading to Eq.~(\ref{cond}), the above expression satisfies the basic
requirement that the density of states scale inversely proportionally
to the atomic volume. The resemblance of Eqs.~(\ref{gamma2}) and
(\ref{g}) is {\em also} notable and, in fact, is far from
coincidental: The quantity $\gamma$ from Eq.~(\ref{gamma2}) can be
thought of as reflecting the strength of intrinsic structural
fluctuations while $g$ reflects the response to an externally imposed
perturbation in the form of a structural defect.

In view of Eq.~(\ref{g}), the interaction between the cooperative
reconfigurations becomes
\begin{equation} \label{Jinf2} J \simeq k_B T_g a^3,
\end{equation}
i.e., comparable to the upper bound on the coupling strength between
the bare degrees of freedom in the Yu-Leggett scenario,
c.f. Eq.~(\ref{Jinf}).  Despite this circumstance, the effects of
interaction (\ref{Jinf2}) are much less important in the presently
discussed local scenario because the local degrees of freedom are
distributed broadly in terms of their energy spacing, the distribution
width being on the order of $k_B T_g$, Eq.~(\ref{nEclass}). Thus the
high value of the laboratory glass transition, relative to the
coupling between the local degrees of freedom, is the reason why it is
difficult for these degrees of freedom to form more than isolated
resonant pairs.

Now, combining Eqs.~(\ref{nEclass}) and (\ref{g}) readily yields an
expression for the quality factor of the phonons:~\cite{LW, LW_RMP}
\begin{equation} \label{llRFOT} \frac{l_\smfp}{\lambda} \simeq
  \left(\frac{\xi}{a}\right)^3 \simeq 10^2.
\end{equation}
We used the numerical value $10^2$ that specifically pertains to the
routine laboratory quench of a molecular substance, as discussed
following Eq.~(\ref{Ntau}). 

Thus we observe that at the purely classical level, the density of
states is set jointly by the glass transition temperature and by the
cooperativity size of the ``mosaic.''  The resulting prediction is
consistent with experiment, Eq.~(\ref{Uratio}). The corresponding heat
capacity and conductivity are then predicted to be, respectively,
strictly linear and quadratic in temperature, at cryogenic conditions.
At the same time, the phonon ``Q-factor'' $l_\smfp/\lambda$ is
determined by the particle content $N^*$ of the cooperative
volume. The latter volume is not strictly universal but depends
logarithmically on the quench speed, by Eq.~(\ref{Ntau}). And so, one
gets generally:
\begin{equation} \label{llRFOTgeneral} \frac{l_\smfp}{\lambda} \simeq
  \left(\frac{\xi}{a}\right)^3 \simeq 16 \left(\frac{d_L}{a}\right)^2
  \ln^2(\tau_g/\tau_0),
\end{equation}
where we used Eq.~(\ref{equip}) and $\tau_g$ is the time scale of the
glass transition. Note that the above result does not explicitly
depend on space dimensionality $d$. The dependence is only implicit,
through the Lindemann ratio $d_L/a$. The formal reason for the lack of
dimensionality dependence is the lack of explicit dependence of the
mismatch penalty term $\gamma N^{1/2}$ on the space
dimensionality. Interestingly, the Lindemann ratio enters into the
above expression in the exactly reciprocal fashion to
Eq.~(\ref{UratioYL}). In view of Eq.~(\ref{adL}), we obtain that {\em
  for a fixed timescale of the glass transition}, the phonon quality
ratio in Eq.~(\ref{llRFOTgeneral}) decreases quadratically with space
dimensionality, that is, not nearly as rapidly as what is prescribed
by the formula (\ref{UratioYL}). Thus in sufficiently many dimensions
and fixed $\tau$, the RFOT-predicted density of states can be very
much lower than the upper limit from Eq.~(\ref{UratioYL}). This is
just another way to see that the emergence of the finite length scale
$\xi$ hinges crucially on the ability of the liquid to reconfigure,
below the crossover to activated transport, on time scales that are
attainable in the sample preparation.

Quantum dynamics does enter into the problem of determining the
density of states explicitly when one estimates the density of
tunnelling states properly so as to account for the ability to carry
out the reconfigurations at low temperatures. To make such an estimate
one must consider the classically defined microcanonical density of
states simultaneously with the distribution of the tunnelling barriers
separating those states, all at a temperature just above $T_g$. In the
early part of the century, Lubchenko and Wolynes~\cite{LW, LW_RMP}
provided detailed calculations for this quantum effect within the RFOT
framework. Here we only point out that the density of configurational
states of our generate aperiodic solid is indeed exponentially large:
$e^{s_c N/k_B}$, thus implying that not only is it possible to find an
alternative state at lengthscale $\xi$, but it is also possible to
find a state reachable from the existing configuration by following an
essentially barrier-less path on a lengthscale that exceeds $\xi$ only
modestly. In volumetric terms, the cooperativity size for such
zero-barrier events is only about 10\% larger than for a
reconfiguration with a typical barrier.~\cite{LW, LW_BP, LW_RMP}
Finding zero-barrier paths requires exploring a somewhat larger region
than the {\em typical} reconfiguration size. The corresponding motions
are thus rare and thermodynamically cannot undermine the mechanical
stability near $T_g$. These tunnelling events through extraordinarily
low barriers require concerted motion of several hundred
particles. The coherence is generally suppressed owing to friction,
i.e., the tunneling event couples to local elastic modes. According to
Refs.~\onlinecite{LW, LW_RMP}, damping becomes particularly important
already at temperatures comparable to the frequency of the
under-barrier motion near the top of the barrier. The top-of-the
barrier frequency turns out to be essentially a fixed fraction of the
Debye frequency $\omega_D$:
\begin{equation} \omega^\ddagger \simeq 1.6 (a/\xi) \omega_D,
\end{equation}
see also Ref.~\onlinecite{LW_Wiley}.

Including quantum-tunneling dynamics in the treatment preserves the
basic scaling of the density of states (\ref{nEclass}) with $T_g$ and
$\xi^3$ but introduces significant multiplicative corrections and
requires use of an adjustable parameter to fit the heat capacity all
the way from cryogenic temperatures to the high temperature set by
$k_B T = \hbar \omega^\ddagger$. Including only the two lowest energy
states in the treatment now yields, up to a multiplicative factor, a
time-dependent heat capacity owing to the distribution of finite
tunneling times. For large times, the theory yields~\cite{LW_RMP}
\begin{equation} \label{power_asympt} \lim_{t\rightarrow \infty} C(t)
  \propto t^{c/2} T^{1+c/2},
\end{equation}
where the constant $c = \hbar \omega^\ddagger/\sqrt{2} T_g$ is
numerically less then $0.1$. The weak time dependence in
Eq.~(\ref{power_asympt}) is consistent with the long time behaviour of
the experimentally observed heat capacity.~\cite{Pohl,
  HunklingerRaychaudhuri, Nittke, Sahling2002} The heat capacity in
the laboratory {\em also} exhibits a mildly superlinear temperature
dependence, but the deviation from strict linearity predicted by this
argument is smaller than what is seen in experiment. We note that the
same approximate arguments~\cite{LW_RMP} that yield the expression
(\ref{power_asympt}) also imply that the heat conductivity is mildly
super-quadratic in temperature. In contast, the experiment seems to
yield a mildly {\em sub}-quadratic dependence.

\begin{figure}[t]
  \includegraphics[width=.75\figurewidth]{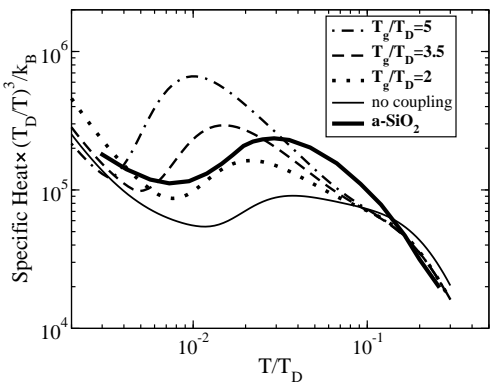}
  \caption{\label{bump} The bump in the amorphous heat capacity,
    divided by $T^3$, as follows from the derived TLS $+$ ripplon
    density of states.~\cite{LW_RMP} The theoretical curves correspond
    to different values of the glass transition temperature relative
    to the Debye temperature, the effects of friction increasing with
    $T_g/T_D$. The thick solid line is experimental data for a-SiO$_2$
    from Ref.~\onlinecite{Pohl}. (For a-SiO$_2$, $k_B T_g/\hbar
    \omega_D \simeq 4.4$.) }
\end{figure}

The problem of the deviation of thermal conductivity from the simplest
prediction is handily resolved, nevertheless, after one realises that
the presence of the lengthscale $\xi$ introduces additional physics
reflected in a new sort of excitation. For one thing, a proper
calculation of the density of states for a region of finite size must
account for zero-point vibrations of its boundary!  Thus an
internally-consistent treatment must include the vibrational
excitations of the domain wall separating cells in the (quantised)
mosaic. We~\cite{LW_BP, LW_RMP} have called these excitations
``ripplons'' to distinguish them from ordinary vibrations of a stable
lattice. Although behaving in many ways like ordinary vibrations---the
intensity of the vibrations goes linearly with $T$---the ripplons are
not simply localised linear vibrational modes of a mechanically-stable
lattice but, instead, correspond to excited states of a structural
resonance whose potential energy surface is bistable or
multistable. (The surface is multi-dimensional of course since $N^*
\simeq 10^2$ particles are involved in reconfigurations.) Despite this
strongly anharmonic aspect of ripplon motions, their frequencies can
be straightforwardly estimated, if one ignores damping effects due to
their interaction with phonons. These modes can be labelled by
spherical harmonic quantum numbers, if we assume the reconfiguring
region is roughly spherical. The mode frequencies are then a simple
combination of the Debye frequency and the dimensionless domain size
$(\xi/a)$:
\begin{eqnarray}
\omega_{l} & \simeq 1.34 & \, \omega_D (a/\xi)^{5/4} \sqrt{(l-1)(l+2)/4}
\nonumber \\
& \simeq & 0.15 \, \omega_D \, \sqrt{(l-1)(l+2)/4}.
\label{om_l_numer}
\end{eqnarray}
Here $l \ge 2$. The second equality is obtained when we use the value
of $\xi$ corresponding to a routine quench as in
Eq.~(\ref{llRFOT}). The curious scaling $(a/\xi)^{5/4}$ is modestly
different from the $(a/\xi)$ scaling expected on geometric grounds
because of the effective curvature dependence of the mismatch penalty
in Eq.~(\ref{FN1}).

\begin{figure}[t]
  \includegraphics[width=.8\figurewidth]{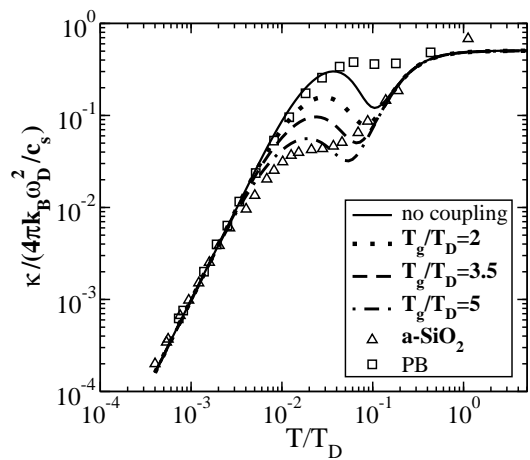}
  \caption{\label{plateau_fig} Heat conductivity~\cite{LW_RMP}
    corresponding with Fig.~\ref{bump}, in comparison with two
    experimental curves.  The (scaled) experimental data are taken
    from Ref.~\onlinecite{smith_thesis} for a-SiO$_2$ ($k_B T_g/\hbar
    \omega_D \simeq 4.4$) and Ref.~\onlinecite{FreemanAnderson} for
    polybutadiene ($k_B T_g/\hbar \omega_D \simeq 2.5$). The low-$T$
    portion is parametrised according to $l_\smfp/l = 150$,
    Ref.~\onlinecite{FreemanAnderson}, while the plateau region is
    computed without using adjustable parameters.}
\end{figure}

The frequencies from Eq.~(\ref{om_l_numer}) fall right within what is
called the Boson peak range. When these excitations are included along
with the two-level excitations, the resulting predictions for the heat
capacity and conductivity match the experiment well, without using any
adjustable parameters, see Figs.~\ref{bump} and \ref{plateau_fig}. The
predicted heat conductivity is not strictly monotonic in the plateau
region, an effect not apparent in experiment. This is likely a result
of ignoring ripplon-ripplon interactions in the theory. A simple
parametrisation of damping effects due to such interaction readily
improves agreement with experiment, but at the expense of using an
adjustable parameter.~\cite{LW_RMP}

We see that a constructive treatment of the anharmonic motions
characterised by an intrinsic length scale $\xi$ of the mosaic
dictates that there must be a significant density of states near the
plateau frequencies.  The resulting description explains in a unified
fashion two sets of seemingly separate cryogenic anomalies: On the one
hand, the approximately linear excess heat capacity and the nearly
universal phonon scattering at sub-Kelvin temperatures are seen as
being caused by the lowest-energy subset of excitations of these local
structural resonances. On the other hand, accounting for the
high-energy, ``ripplonic'' part of the spectrum of the local
resonances already at the simplest level quantitatively accounts for
the excess specific heat and phonon scattering at the plateau
temperatures. Lubchenko and Wolynes, Section V of
Ref.~\onlinecite{LW_RMP}, have outlined steps for a more proper way to
quantise the mosaic, so as to account for both the repulsion between
the classically obtained energy levels and the effective
renormalization of individual tunnelling amplitudes due to the rest of
the tunnelling motions. That discussion indicates that the diagonal
and off-diagonal elements of the tunnelling centres, such as the
quantities $\epsilon$ and $\Delta$ from Eq.~(\ref{HTLS}), are in fact
correlated, in contrast with standard treatments.  The resulting
distribution~\cite{LW_RMP} of $\epsilon$ and $\Delta$ is no longer
factorisable; it now has two relatively distinct contributions. One
contribution comes from highly quantum TLS, whose classical energy
splitting $\epsilon$ is effectively pinned near zero. The other
contribution comes from those TLS that more closely resemble the
prescription of the standard tunnelling model (STM). This predicted
dichotomy between highly quantum and quasi-classical structural
resonances would seem to explain the apparent existence of ``fast''
and ``slow'' tunnelling systems noted early on by Black and
Halperin.~\cite{BlackHalperin} The ``fast'' two-level systems would
account for the surprisingly fast onset of heat capacity---faster than
that predicted by the STM---seen in experiment. Also in an improvement
over the results from the purely quasi-classical analysis, the density
of states---once corrected for the effects of level repulsion and the
renormalisation of tunnelling amplitudes---becomes weighted more
heavily toward higher energy, consistent with the mildly sub-quadratic
temperature dependence of the heat conductivity seen in the
laboratory.

A rather exotic yet consequential manifestation of the ripplons is
that, together with the underlying structural transition, they are
predicted~\cite{LW_RMP} to quantitatively account for the negative
thermal expansivity seen in many glasses.~\cite{Ackerman} The idea is
that the phonon-mediated dispersion forces between thermally active
local resonances effectively squeeze the sample. This squeezing is
entirely analogous to the Casimir effect between plates in a vacuum,
but with the phonons filling in for the photons. The quantity of such
thermally active local resonances grows sufficiently fast with
temperature so as to counteract the intrinsic propensity of solids to
expand with temperature owing to the asymmetry of near-neighbour
inter-particle interactions. The present author is not aware of any
attempts to rationalise the unusual negative thermal expansivity
within any other theoretical frameworks. Although a negative
Gr\"{u}neisen parameter~\cite{Kittel} is not uncommon in complex
crystalline solids, its common appearance in amorphous
materials---which are isotropic down to very short lengths---seems
rather surprising. In my view, the natural emergence of a negative
thermal expansivity in this picture lends it substantial support.

Additional support for the quantum mosaic picture came relatively
recently from a somewhat unexpected direction: Zhugayevych and
Lubchenko have argued~\cite{ZL_JCP} that in some substances with
correct atomic orbital structure, the domain walls separating distinct
aperiodic minima should host special, very deep midgap electronic
states analogous to the solitonic midgap states in
trans-polyacetylene.~\cite{RevModPhys.60.781} Specific, quasi-one
dimensional molecular motifs were generated where the midgap state is
centred around an over-coordinated or under-coordinated
atom.~\cite{ZLMicro2} More generally, such midgap states are
expected~\cite{} to reside on interfaces between distinct states of
charge-density waves.~\cite{GHL} The deep midgap states have many
interesting properties: They display a reverse charge-spin relation. A
half-filled state, which has spin 1/2, is electrically neutral while
the filled and empty states are, respectively, negatively and
positively charged but nevertheless spinless. Exactly such states
appear to be present in amorphous chalcogenides: Pristine samples
exhibit a relatively clean absorption edge and show no detectable ESR
signal. Yet after being exposed to macroscopic quantities of photons
at near-gap frequencies, these chalcogenides manifest a simultaneous
increase in midgap absorption and number of unpaired
spins.~\cite{BiegelsenStreet, PhysRevB.38.11048} Both the number of
midgap absorbers and the number of unpaired spins seem to saturate at
$\sim 10^{20}$~cm$^{-3}$. This density is in remarkable agreement with
the RFOT theory-based estimate for spatial concentration of the domain
wall regions, Eq.~(\ref{xiconc}). The figure $\sim 10^{20}$~cm$^{-3}$
is interesting in that it is significantly greater than the typical
concentration of dopants in crystalline semiconductors. It is
remarkable that a material could host such large quantities of what
seems to be intrinsic ``defects''---which become apparent upon
irradiation---while not being amenable to conventional
doping.~\cite{Kolomiets1981}

\begin{figure}[t] \centering
  \includegraphics[width= \figurewidth]{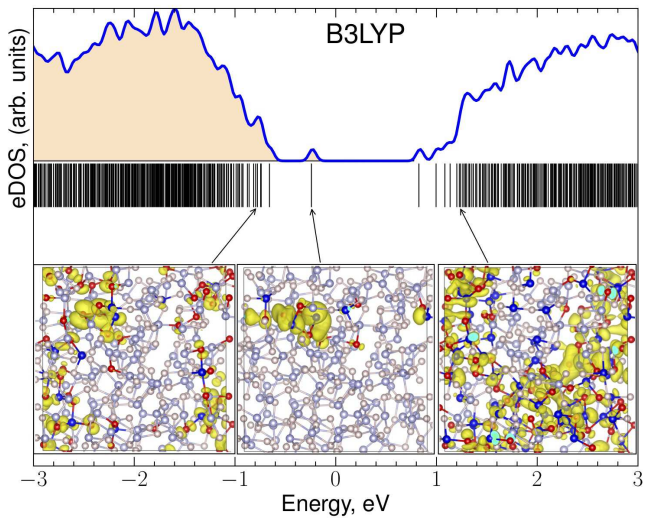}
  \caption{\label{midgap} {\em Top}: The electronic density of states
    of an amorphous sample of As$_2$Se$_3$ obtained in
    Ref.~\onlinecite{LL2} using the structure-building algorithm
    developed in Ref.~\onlinecite{LL1}. A very deep midgap state is
    visible, whose wavefunction is shown in the bottom panel,
    alongside the wavefunctions of two select states near the edges of
    the two mobility bands. The acronym ``B3LYP'' in the top panel
    signifies the specific quantum-chemical
    approximation~\cite{vasp1,vasp2,vasp3,vasp4} used to obtain the
    electronic spectra. Image courtesy of Dr. A. Lukyanov.}
\end{figure}

Already in 1975, Anderson~\cite{PWA_negU, PWA_negU2} proposed his
negative-$U$ model, which postulates effective attraction between
electrons residing on a relatively localised orbital. The model
provides an elegant mechanism for efficient pinning of the Fermi level
near the middle of the forbidden gap, viz., at the energy where a {\em
  singly} occupied centre would be. Because of the negative Hubbard's
$U$, the defect states are typically filled and are optically active
but at supra-gap frequencies. Anderson speculated already in 1975 that
the putative negative-$U$ physics may be related to the two-level
systems, a prescient insight indeed.  Yet the interactions responsible
for the effective electron-electron attraction on a particular site
would be present at {\em every} site. As such, these interactions
would also determine the stabilization due to forming a filled valence
band and hence the value of the forbidden gap
itself.~\cite{ISI:A1972L738100004} This circumstance makes the
applicability of the negative-$U$ scenario to chalcogenides
questionable. The gap in the chalcogenides, crystalline or amorphous,
is largely due to the charge density wave arising from the spatial
variation in both bond strength and local
electronegativity.~\cite{ZLMicro1, GHL} In contrast to the original
negative-$U$ proposal, the midgap states in the picture advanced by
Zhugayevych and Lubchenko are robust for topological
reasons.~\cite{PhysRevD.13.3398} This robustness comes about because
malcoordination cannot be removed by small lattice deformations, but
only by breaking/creating a bond or recombination of opposite
malcoordinations. Like in the original negative-$U$ model scenario,
there is still some stabilisation of a filled state, however the
strength of stabilisation is not directly linked to the gross chemical
interactions that determine the gap size. Instead, the stability has
to do with relatively subtle polarisation effects occurring when an
electron (hole) is added to a radical. Several specific bonding
configurations can be argued to be stabilised by the addition of
electrons/holes to neutral defects already using simple ideas \`a la
G.~N.~Lewis.~\cite{ZLMicro2} Recently, {\em bulk} samples of amorphous
chalcogenides exhibiting such midgap states have been
computationally-generated.~\cite{LL1} According to
Ref.~\onlinecite{LL2}, not only do these computationally generated
samples appear to host both the mobility-band and the associated
Urbach tail~\cite{PhysRev.92.1324, CFO, PhysRevLett.57.1777,
  Kostadinov} of localised states, but they also exhibit just the type
of topological states that were predicted by the ZL picture.  The
electronic spectrum of the sample and the wavefunction of the midgap
state are exemplified in Fig.~\ref{midgap}.

Finally we point out that at temperatures below but not too much below
the crossover to the activated transport, the cooperative
reconfigurations are less compact and are supplemented by a different
type of motion that is locally string-like.~\cite{SSW} These motions
correspond to transient opening of the local cages and have been
argued to make a universal contribution to the so called
Johari-Goldstein,~\cite{JohariGoldstein} or
$\beta$-relaxation.~\cite{SWbeta} From the string viewpoint, the
transition from the aperiodic solid $\alpha>0$ to the uniform liquid
$\alpha=0$ can be thought of as a string-deconfinement transition, in
which string-like motions percolate ultimately resulting in molecular
translations becoming barrier-less;~\cite{SSW} we will return to this
important notion shortly. We note that other, system-dependent local
motions can be present in glassy materials, such as conversions
between rotamers in branched polymers.

\section{The global marginal-stability scenario}

\label{marginal}

\begin{figure}[t]
  \includegraphics[width= 0.9 \figurewidth]{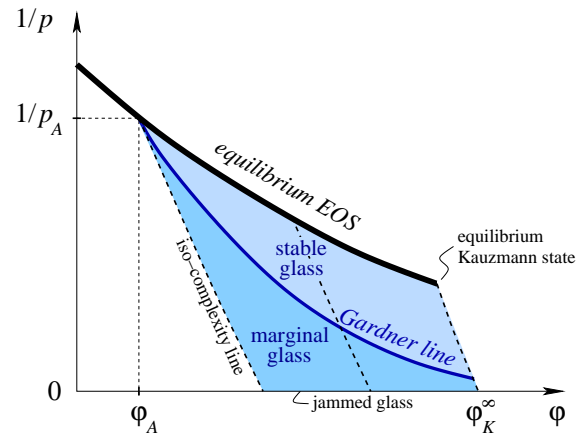}
  \caption{\label{MF} The phase diagram of a meanfield liquid in the
    $(\varphi, p^{-1})$ plane, after
    Ref.~\onlinecite{Berthier26072016} The quantities $\varphi =
    \rho/(\pi \sigma^3/6)$ and $p$ are the filling fraction and
    pressure respectively, $\sigma$ the diameter of the
    sphere. EOS~$=$~equation of state. Note reports on the location of
    the low-$\varphi$ end of the Gardner line
    vary.~\cite{Charbonneau2014}}
\end{figure}

A rather distinct-looking explanation for low-$T$ universalities in
glasses has grown up in recent years within strictly meanfield,
replica symmetry breaking (RSB) treatments of the structural glass
transition. In a fascinating set of
developments,~\cite{doi:10.1021/jp402235d, Charbonneau2014,
  Mariani24022015, Berthier26072016, PhysRevLett.114.125504} the
latter treatments indicate that for a sufficiently deep quench within
a specific free energy minimum, the local motions within their cages
themselves would eventually exhibit marginal stability and,
subsequently, will undergo further symmetry breaking in the form of a
continuous phase transition. This transition is formally analogous to
the infinite step RSB found in the Sherrington-Kirkpatrick
model~\cite{PhysRevLett.35.1792} and also more pertinently to the
Gardner transition in $p$-spin models.~\cite{Gardner1985747,
  GrossKanterSomp} To make direct connections among the problems of
the glass transition, jamming, and a variety of packing
problems,~\cite{RevModPhys.82.789} it is often convenient to consider
not only thermal quenches at constant pressure, but also pressure
quenches at constant temperature. The two types of quenches are
straightforwardly related for strictly rigid particles, whose equation
of state contains the pressure and temperature in the combination
$p/T$. For rigid spheres, the density and volume fraction can be used
interchangeably.

To describe the Gardner transition, we first recall that
meanfield-theory liquids---similarly to Potts spin glass and $p$-spin
models---exhibit a one step replica symmetry breaking at some
temperature $T_A$ (or pressure $p_A$), which corresponds with the
emergence of a thermodynamic quantity, $e^{s_c N/k_B}$, of equivalent
free energy minima. In meanfield theory, each minimum becomes
separated from the rest of the minima by strictly infinite barriers;
the structure corresponding to each minimum is aperiodic. The
transition signifying the emergence of this complex free energy
landscape is depicted by the $(\varphi_A, 1/p_A)$ point in
Fig.~\ref{MF}.

Suppose now one quenches a liquid starting from a state with $\alpha >
0$, i.e., when a liquid is already securely arrested in one of those
equivalent, infinitely-deep aperiodic free energy minima. As the
quench continues further, a Gardner transition occurs
eventually,~\cite{doi:10.1021/jp402235d} where the current free energy
minimum splits into two basins. The distinct free energy sub-basins
that emerge in this way now have a distributed degree of similarity;
the replica symmetry breaking is now
continuous.~\cite{SpinGlassBeyond} The instability leading to this
continuous RSB points to a rather subtle {\em vibrational} property of
(high-dimensional) aperiodic solids whose detailed nature is not
entirely understood at present. Here we point out that the continuous
replica symmetry breaking (RSB) implies that there is an additional
contribution to the system's entropy, in the form of the mixing
entropy of the distinct replicas. The corresponding degree of freedom
is an exchange of particles between
replicas.

The appearance of marginally stable modes on the approach to the
Gardner instability, following a deep quench, may have intriguing
microscopic implications. Indeed, since the Gardner transition is a
critical point, it has been conjectured that the resulting scale
invariance could potentially explain the universality of the cryogenic
anomalies this article has focused on. In seeming support of this
suggestion, Gardner-like instabilities accompanied by long correlation
lengths are observed in simulations of polydisperse hard sphere
liquids, where particle-swap moves are explicitly
allowed.~\cite{Berthier26072016} On the other hand, analogous
simulations of polydisperse {\em soft} spheres to not seem to show
long range correlations, while the instabilities are rather localised
and only sporadic.~\cite{2017arXiv170604112S}

\begin{figure}[t]
  \includegraphics[width=  \figurewidth]{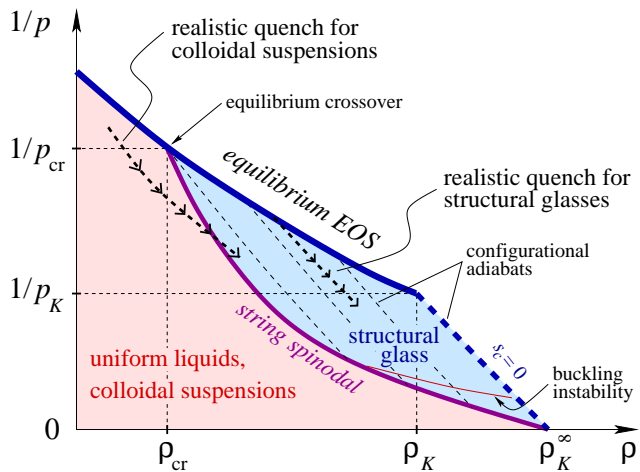}
  \caption{\label{nonMF} The off-equilibrium diagram of a
    non-meanfield liquid in the $(\rho, p^{-1})$ plane, where $\rho$
    is the number density. $s_c$ stands for the configurational
    entropy. Liquid states cannot exist to the right of the $s_c=0$
    line, by construction. Note that the ``string spinodal'' is a
    crossover, not a sharp boundary, in view of the limited spatial
    extent of aperiodic free energy minima in finite dimensions.  }
\end{figure}

Recent arguments by Lubchenko and Wolynes~\cite{LWjamming} (LW), which
go beyond meanfield theory, nevertheless indicate that the
implications of the Gardner scenario for actual glass-forming
substances made of atoms are only limited. (To avoid confusion, we
note that amorphous solids comprise a broad class of assemblies of
particles ranging from chemical substances to jammed colloidal
suspension to heaps of grains.) Some of the results of LW's
non-meanfield analysis are graphically summarised in
Fig.~\ref{nonMF}. In fundamental distinction from meanfield
approaches, in which relaxation barriers are infinite by construction,
LW~\cite{LWjamming} discuss quenches of particulate assemblies at a
finite rate. Two distinct physical situations are ordinarily realised
depending on the magnitude of the dimensionless {\em kinetic}
pressure, $p/k_B T \rho$ of the particle assembly. For molecular
glassformers, this kinetic pressure is about four orders of magnitude
larger than the ambient pressure,~\cite{L_AP} being numerically in the
Gigapascals or higher. Defining such a kinetic pressure relies on the
particles' possessing an effective rigid core. This is an excellent
approximation for atoms at normal pressures, which do exhibit a
relatively rigid ionic core. This is because exciting electrons out of
the the core requires energies much exceeding the thermal energy.  The
kinetic pressure is nearly completely counterbalanced by the cohesive
interactions between the atoms; the difference is comparable to the
atmospheric pressure, of course.  Now, the time interval between
inter-particle collisions is very short in such systems, significantly
less than a picosecond and so vibrational equilibration is very
quick. Thus even if subjected to the fastest quench realisable in a
laboratory, such a system will undergo the translational symmetry
breaking $(\alpha = 0) \to (\alpha \approx 10^2)$ while still
equilibrated. {\em Further} quenching results in a rapid increase in
the relaxation times, as discussed in Section~\ref{local}. Once the
quench rate matches the rate of the slowest motions in the liquid, a
kinetic glass transition takes place; the system falls out of
equilibrium and now follows a non-equilibrium equation of state, in
which the configurational entropy is approximately steady: $s_c =
\text{const}$. These off-equilibrium equations of state are indicated
by thin dashed lines in the high-density, blue sector on the diagram
in Fig.~\ref{nonMF}. After falling out of equilibrium, the quench will
proceed approximately along a configurational adiabat $s_c =
\text{const} > 0$.

Now suppose the effective rigid core of the particles does not
significantly decrease in size with pressure (as it actually does for
real molecules!). In that case, the quenched state can become
arbitrarily higher in Gibbs free energy than does the closest packing
of the particles: $\Delta G = \int V dp$. The expression (\ref{FN1})
for the free energy cost of reconfigurations can be generalised for
situations where the initial state is off-equilibrium so that the
driving force now includes an enthalpic component.~\cite{LW_aging,
  Wisitsorasak02102012, LWjamming} The resulting relaxation represents
{\em ageing}. At high pressures, the total driving force per unit
volume, $\approx \Delta G/V$, scaling approximately with the external
pressure, can thus be made arbitrarily high in the $p \to \infty$
limit.  A detailed calculation~\cite{LWjamming} shows that despite the
mismatch penalty {\em also} increasing with pressure, the
cooperativity size for relaxation can become arbitrarily small as the
pressure increased, while the corresponding reconfiguration barrier
saturates at a fixed value. Physically this means that the system
becomes unstable with respect to one-particle ageing events that
destroy the cage that had been formed by the particle's nearest
neighbours as a result of the breaking of the translational symmetry
during the formation of the aperiodic crystal.

\begin{figure}[t]
  \includegraphics[width= .9 \figurewidth]{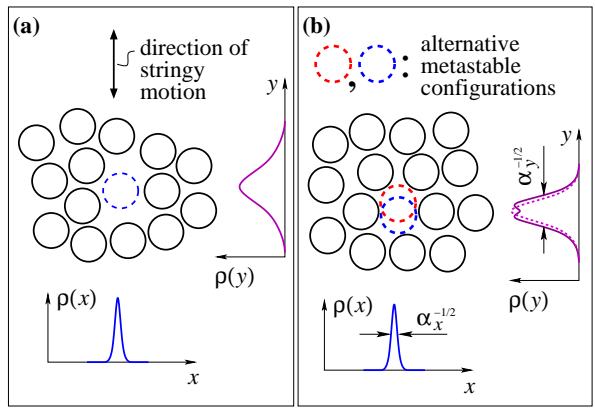}
  \caption{\label{jamstrings} {\bf (a)} A stringy instability is
    signalled by the ability of a particle to escape the cage via a
    displacement that significantly exceeds the vibrational
    magnitude. The vibrational amplitude in the direction of lower
    pressure exceeds its typical value, while the collision frequency
    is lowered.  {\bf (b)}~Illustration of how compression of an
    anisotropic cage can lead to a local symmetry breaking upon which
    the caged particle will be subject to a bistable effective
    potential and will have to choose one of the minima upon further
    compression, while leaving the remainder of the cage slightly
    undercoordinated. }
\end{figure}

The cage-escape events are locally string-like, see
Fig.~\ref{jamstrings}(a). The strings will generally emanate from
multiple locations and percolate, thus resulting in zero-barrier
relaxation events and partial restoration of translational
symmetry. These notions are entirely analogous to the microscopic
picture put forth some time ago by Stevenson, Schmalian, and
Wolynes~\cite{SSW} in the context of the {\em equilibrium} crossover
between activated and collisional regimes in finite dimensions. The
reader is reminded that the position of this equilibrium crossover is
much lowered relative to the meanfield spinodal at $T_A$ ($p_A$) where
metastable minima in the liquid just begin to form but are easily
overcome by thermal fluctuations.~\cite{LW_soft} LW
argued~\cite{LWjamming} that a similar instability would take place in
quenched glasses as well, with the difference that relaxation is now
also downhill in the free energy of an {\em individual} aperiodic
minimum and thus results in structures that are significantly
stabilised enthalpy-wise.

A separate argument~\cite{LWjamming}---which is related to the string
argument, but is somewhat different in spirit---suggests that a milder
instability can occur in systems of particles at pressures that are
not quite as high as those required for the string-instability to
occur. Here one notices that owing to the inherent lack of symmetry in
the immediate coordination shell of a particle in an aperiodic
lattice, the cages formed around each particle are anisotropic. Thus
sufficiently strong compression can lead to a situation where for a
particle located close the centre of its cage, the number of contacts
it makes with its immediate neighbours could decrease below the
minimum number of constraints needed for mechanical stability, see
Fig.~\ref{jamstrings}(b). According to
Maxwell,~\cite{MaxwellConstraint} this number is equal to $2d$ in $d$
spatial dimensions. Accordingly, this unstable configuration must
separate two configurations each of which may be metastable. Further
compression will result in the particle having to choose one of the
metastable configurations, while the vacant end of the original cage
will partially collapse, or ``buckle.'' Instabilities of a similar
type have been also discussed by Wyart and
coworkers.~\cite{PhysRevE.72.051306, doi:10.1063/1.3157261,
  DeGiuli02122014}

The emergence of two new metastable configurations from one is a
symmetry breaking. Because this symmetry breaking results in further
stratification of the free energy landscape that emerged at the
equilibrium crossover, it can perhaps be thought of as a finite
dimensional realisation of the Gardner transition. Whether this is
case is not clear since buckling could cause further relaxations;
complete analysis of this possibility is presently lacking. In any
event, for the buckling instability to take place, the difference
between the particle displacements along the long and short axes of
the cages must exceed the amount a bond can deform under the pressure
in question. This notion may be relevant to the aforementioned lack of
divergent correlations in simulations of soft
particles.~\cite{2017arXiv170604112S} In the context of actual
glassformers, one may further conclude that buckling would be
efficiently suppressed by quantum effects.

Thus at a first glance, the string-spinodal and the buckling
instabilities represent candidates for the emergence of marginal
stability in finite-dimensional systems with potential implications
for the emergence of anharmonic degrees of freedom in cryogenic
glasses. Yet already for rigid spheres, qualitative
estimates~\cite{LWjamming} indicate that to reach these instabilities,
one must achieve kinetic pressures that are {\em multiples} of the
kinetic pressure the atoms exert on each other at normal
conditions. Ordinary, thermally-quenched glasses are nowhere close to
this regime, of course.  In addition, any incipient criticality due to
such instabilities would be destroyed by the activated reconfiguration
events that occur by ageing because at a critical point, relaxation
times strictly diverge~\cite{Goldenfeld} whereas the activation time
remains finite and, in fact, decreases with pressure on approach to
the string spinodal.~\cite{LWjamming} This is entirely analogous to
the way the criticality due to mode-mode coupling kinetic catastrophe
at the meanfield transition at temperature $T_A$ is destroyed by the
activated transitions.~\cite{LW_soft, MCT, KTW}

It seems instructive to make an explicit connection between the LW
arguments and Yu-Leggett philosophy. It is possible to estimate the
coupling between the putative Gardner degrees of freedom exemplified
in Fig.~\ref{jamstrings}(b).  For the sake of concreteness, we will
work with rigid particles implying enthalpy changes come from pressure
variation, at constant temperature: $dH = Vdp$.  The amount of lattice
distortion along the $y$-axes gained over the displacement
$\alpha^{-1/2}_y$, relative to the typical in-cage displacement
$\alpha^{-1/2}_x$, is equal to $(\alpha^{-1/2}_y/\alpha^{-1/2}_x)
\nabla \phi$. The resulting enthalpy bias gives the coupling of the
transition to the displacement field: $g \simeq pa
(\alpha_x/\alpha_y)^{1/2}$. Further recalling that the bulk modulus
for rigid particles goes as $K/k_ B T\rho \simeq (p/k_B T
\rho)^2$,~\cite{LWjamming} we obtain:
\begin{equation} \label{gmarginal} g \simeq (\alpha_x/\alpha_y)^{1/2}
  (K k_B T/\rho)^{1/2},
\end{equation}
consistent with Eq.~(\ref{g}).  Indeed, during string-like motions
that occur near the crossover, particle displacement in one direction
significantly exceeds displacements along the other directions. This
can be formally viewed as the ratio $(\alpha_x/\alpha_y)^{1/2}$
greatly exceeding 1. Such degrees of freedom would interact with each
other much more strongly than do the modes in a bona fide aperiodic
crystal, in which $\alpha_x \approx \alpha_y$, thus resulting in their
ready freezing-out.

We observe that since for the Gardner modes, $\alpha_y^{-1/2}$ does
not exceed $\alpha_x^{-1/2}$ by much, their energetics are similar to
the energetics of the local structural reconfigurations reviewed in
Section~\ref{local}. The basic energy scale for the density of states
is determined by the only pertinent energy scale in the problem, viz.,
the pressure at the Gardner transition, times the particle
volume. This is numerically similar to the temperature at the
beginning of the quench and thus is quite analogous to the local
scenario from Section~\ref{local}.  Likewise, the infinite-range
mechanism of the universality from Section~\ref{infrange} will not
apply here because of the lack of resonance between the Gardner modes.

Finally we briefly comment on the amorphous solids made by quenching a
liquid starting from a configuration on the low-density side of the
string-spinodal in Fig.~\ref{nonMF}. Such quenches would be typical
for colloidal suspensions whose ``molecular'' time-scale $\tau_0$ is
much much larger, easily by ten orders of magnitude, than that in
molecular liquids proper. This is because the microscopic time scale
$\tau_0$ is now determined by the solvent's viscosity and the (very
large) particle size. Lubchenko and Wolynes~\cite{LWjamming} argued
that rigidity emerges in such colloidal-particle assemblies following
the crossing of the string-spinodal at a point that is far away from
equilibrium. Under these circumstances, local regions will undergo
zero-barrier relaxation events leading to the emergence of activated
dynamics in those regions, possibly followed by avalanches. Because of
the large value of $\tau_0$, even the smallest barrier will result in
the relaxed region being virtually rigid. The rigid regions formed in
this way will percolate resulting in a jammed solid. Although very
interesting in their own right, such jammed solids are essentially
classical, because of the large particle mass. This large mass would
altogether prevent observing the presently discussed quantum anomalies
in jammed solids.

\section{Summary and experimental tests going forward}

\label{summary}

We have reviewed a number of ways to think about the
quantum-mechanical freezing out of the liquid degrees of freedom in
glassy solids as they approach absolute zero. These ``relics'' of
liquid motions amount to a substantial density of states in excess of
the Debye-like vibrational degrees of freedom that reveals itself in
the form of resonances at sub-Debye frequencies.

The Yu-Leggett scenario, Section~\ref{infrange}, postulates the
existence of some bare degrees of freedom that are not explicitly
specified. Despite this agnosticism, one may generate a criterion for
mechanical stability with respect to compound excitations involving
motions of several bare degrees of freedom that interact.  This
analysis then yields a meaningfully restrictive upper bound on the
density of structural states is excess of the vibrational motions of a
fully stable solid, at least in the physical three dimensions. These
excess structural excitations are viewed as resonances formed by the
bare degrees of freedom; the spatial extent of the resonances
resonances can be arbitrarily large, yet it is lower than the
wavelength of thermal phonons thus enabling these compound degrees of
freedom to scatter phonons resonantly.  There appear to be two
distinct scattering regimes. At low temperatures, where the
thermally-pertinent interaction range between the bare degrees of
freedom significantly exceeds the molecular length scale, phonons have
a relatively high-fidelity, $l_\smfp/\lambda \gtrsim 10^2$. As the
temperature is raised, the system crosses over to a regime
characterized by complete phonon localization, $l_\smfp/\lambda
\gtrsim 1$. We saw that for the spatial dimensionality greater than
the physical value of 3, even the low-$T$ bound on the
$l_\smfp/\lambda$ ratio quickly drops to the Yoffe-Riegel value of one
thus limiting the usefulness of this bound for estimating the actual
amount of phonon scattering.

The scenario advanced in Section~\ref{local}, using the RFOT theory,
is more concrete about the origin of two-level systems. In that
approach, the lifetimes of metastable, transiently-rigid local
structures are explicitly computed, alongside the corresponding
reconfiguration size. The experimentally observed, gradual onset of
rigidity in the classical regime is quantitatively reproduced. This
gradual onset of rigidity is associated with the growth of the size
$\xi$ of mechanically stable regions.  The value of this cooperativity
size $\xi$ near the glass transition---where the liquid falls out of
equilibrium---along with the glass transition temperature itself, then
determine the gross density of states of the residual liquid motions
that have not frozen out as a result of the glass transition but that
now must equilibrate quantum mechanically at low temperatures.
Further quantum effects have been analysed and lead to improvements in
the estimate of the density of states and time dependence of the
specific heat. In the LW scenario, the phonon ``quality'' ratio
$l_\smfp/\lambda$ is determined by the size of the cooperative region
$(\xi/a)^3$, which is predicted by the RFOT theory; it depends on the
quench speed weakly, logarithmically, thus pegging it near the value
of order $10^2$ for glasses made by conventional quenches. The finite,
even if relatively modest, lengthscale $\xi$ characterises the
lengthscale of a ``mosaic'' comprised of distinct aperiodic solutions
of the free energy functional. The domain walls separating the mosaic
cells are relatively strained regions that give rise to qualitatively
new physics that are not directly accessible to meanfield
approximations. In particular, vibrational excitations of the domain
walls quantitatively account for the phenomena associated with the
Boson peak and the thermal conductivity plateau. In amorphous
semiconductors characterised by spatial variation of bond saturation,
the domain walls are also predicted to host special, very deep midgap
electronic states which are revealed by exposing a pristine sample to
macroscopic amounts of photons at near gap frequencies.

The non-meanfield nature of the RFOT-based calculation is key to one's
ability to capture physics associated with reconfigurations on {\em
  finite} timescales. The finiteness of the cooperativity length $\xi$
is directly linked with the finiteness of the relaxation barrier in a
liquid near the glass transition, which is kinetically controlled. The
present author believes that the mechanical stability of a glassy
liquid on lengths less than $\xi$ can be thought of as the bare
degrees of freedom from the Yu-Leggett infinite-range scenario largely
frozen out on all length scales less than $\xi$. As a result, these
degrees of freedom no longer contribute significantly to phonon
scattering while the cooperative motions on the length $\xi$ are the
only remnants of the liquid motions.  The density of states of these
nanoscopic regions is considerably lowered in comparison with that for
the bare degrees in the infinite-range scenario due to Yu and
Leggett. Indeed, the typical energy spacing is now on the order of
$T_g$ implying resonant interactions between these degrees of freedom
are unlikely and do not significantly affect thermal properties.  The
interaction still manifests itself in subtle ways already at
near-Kelvin temperatures because of their multilevel nature, for
instance, by causing a negative thermal expansivity.  Direct
determinations of the cooperativity length $\xi$, together with dozens
of quantitative predictions made by the RFOT theory~\cite{LW_ARPC,
  L_AP}---including predictions for the density of states of the
excess structural degrees of freedom observed in cryogenic
glasses---give confidence in this microscopic picture. Some of the
most direct microscopic signatures of the non-trivial length $\xi$ in
frozen glasses are the Boson peak and the topological midgap
electronic states in amorphous semiconductors.

The ability of a glassy liquid to stabilise or {\em age} via
cooperative reconfigurations may be also key for assessing the
relatively recent proposal for the low-$T$ glassy anomalies as
stemming from marginal stability on the approach to Gardner
instabilities predicted to take place in meanfield liquids. In the
latter picture, discussed in Section~\ref{marginal}, the liquid is
unable to reconfigure below the temperature $T_A$ at which metastable
structures begin to form and can be thought of as residing all in the
very same aperiodic free energy minimum. This is because
interconversions between the minima are subject to strictly infinite
barriers in the meanfield limit.  Yet in finite dimensions the liquid
is broken up into a mosaic made of distinct aperiodic solutions, as
discussed in Section~\ref{local}. The cost of making the mosaic,
$\gamma N^{1/2}$ per domain, is exactly compensated by the entropic
stabilisation, $-Ts_c N$ per domain, at the cooperative size
$N^*$. Still, the mosaic is more stable than a bulk aperiodic state,
at finite temperatures, because of the finite entropy of the domain
walls.~\cite{EastwoodW} The Gardner-unstable regions---where the free
energy density is typically higher---thus can be viewed as resulting
at least in part from the inability of the liquid to fully
(entropically) stabilise.


Going beyond meanfield theory, Lubchenko and Wolynes~\cite{LWjamming}
have argued that quench-induced instabilities in finite dimensions
would require very high kinetic pressures that are in significant
excess of those achievable in a glass using a conventional thermal
quench. In fact, achieving such kinetic pressures requires ambient
pressures on the order of several Gigapascales. In addition, any
incipient criticality associated with the Gardner instability would be
destroyed by the activated ageing events. For these reasons, the
Gardner instabilities were argued not to be the cause of the cryogenic
anomalies observed in structural glasses.

On the question of scale invariance or absence thereof in the context
of the emergence of rigidity in structural glasses, one may
legitimately enquire as to the current status of renormalisation group
(RG) based treatments of the glass transition. Such treatments have
often been the tool of choice in testing for scale invariance. As
pointed out in Ref.~\onlinecite{Lfutile}, RG calculations are
particularly difficult to implement for liquid-to-solid transitions
because the very nature of the degrees of freedom changes entirely at
the transition, viz., from density fluctuations above $T_\scr$ to
vibrations of an essentially fixed lattice below $T_\scr$. These
difficulties can be partially circumvented in
glasses~\cite{AngeliniBiroli} by using a related spin model that
happens to exhibit, in the meanfield limit, the dynamical transition
analogous to the crossover and even the putative Kauzmann singularity
at the temperature $T_K$, Eq.~(\ref{scT}). If one could equilibrate
the sample at $T_K$, the configurational entropy would vanish while
the relaxation barrier would strictly diverge.

Angelini and Biroli~\cite{AngeliniBiroli} have argued, using a spin
model and an approximate RG scheme that above spatial dimensions 4,
one indeed should expect two zero temperature fixed points that
correspond with the crossover and the Kauzmann crisis. ``Zero
temperature'' implies that the ratio of the coupling strength to
temperature diverges at the transition. At finite temperatures, this
corresponds with an infinitely deep bound state. Such bound states are
characteristic of asymptotically free
theories.~\cite{RevModPhys.77.837, RevModPhys.77.857} In the context
of solids, such infinitely deep bound states are natural: To break a
single bond, an infinitely many bonds must be broken. Incidentally,
Bevzenko and Lubchenko~\cite{BL_6Spin, BLelast} have shown at the
Onsager-cavity level that elasticity is, in fact, an asymptotically
free theory. For dimensions 4 and below, however, those fixed points
are avoided. In 3D, the correlation length reaches a value numerically
close to 15, instead of diverging. In view of Eq.~(\ref{xia6}), even
such a modest size implies super-cosmological relaxations
times,~\cite{Lfutile} thus making it impossible to observe avoided
criticality in conventional glass-formers.  We must also note that
even if the Kauzmann state does in principle exist in
3D---extrapolations of the configurational entropy suggest that it
does!---detailed calculations still indicate that the Kauzmann state
would be avoided in 3D owing to nano-crystallisation or a similar type
of local ordering.~\cite{SWultimateFate}

One of the challenges posed by the cryogenic anomalies in glasses is
that direct probes of the spatial extent of the structural resonances
have not been found in the laboratory. We reiterate that there are, in
principle, three relatively distinct options for the resonance size:
One is that these puzzling degrees of freedom involve motions of at
most a few particles, as in the original TLS proposal~\cite{AHV,
  Phillips} or its soft-potential generalisations.~\cite{soft_pot,
  soft_potBuchenau, Klinger} Another size follows from the RFOT
theory, which dictates that while the resonances are still local---and
quite compact!---they involve significantly more atoms, several
hundred or so, and thus each span a region a few nanometers
across. Finally in infinite-range scenarios \`a la Yu and Leggett, the
resonance could be formed by two primitive degrees of freedom that
could be separated by an arbitrarily large distance. Since even such
arbitrarily extended degrees of freedom can absorb/emit phonons
resonantly at the low energies in question, existing experiments that
probe defect states, such as phonon echo~\cite{GG} or spectral
diffusion,~\cite{KlauderAnderson, BlackHalperin, LS_EnMass} do not
probe the spatial extent of those defects. The situation is even worse
when the chromophore is coupled to the structural resonances
electromagnetically since for the same amount of exchanged energy, the
wavelength of light is much greater than that for sound. Incidentally,
the apparent dipole moment for structural rearrangements is relatively
large, on the order of a Debye.~\cite{Maier} Within the RFOT-based
framework,~\cite{LSWdipole, Lionic} such a large dipole moment comes
about through a cumulative effect of a large number of small elemental
dipole moments generated by rotations of a close pair of particles in
a polar substance. Conversely, it is difficult to imagine how such a
large dipole moment would be generated by motion of just a few
particles in a very dense glass. I note in passing that the
``electrodynamics'' of the two-level systems is not limited to their
exhibiting a dipole moment. Some of the glasses show very puzzling
magnetic-field effects,~\cite{PhysRevLett.88.075501} discussion of
which is beyond the scope of this article; a recent review is
available in Ref.~\cite{doi:10.1080/14786435.2015.1109717} It appears
that these magnetic field effects are consistent with the
multi-particle picture advanced by the RFOT theory.

To address some of the challenges faced by existing spectral diffusion
setups, in which one looks at a single chromophore at a time,
Lubchenko and Silbey proposed~\cite{LSbinoculars} that such setups can
be modified to separately determine the concentration of the
resonances and their coupling to the phonons (or photons). In this
modified setup, one employs not one but two or more chromophores and
monitors simultaneously the evolution of their frequencies. The idea
behind such ``molecular binoculars'' is that high-frequency,
low-magnitude spectral jumps of the chromophores are due to remote
defects. Such jumps will be correlated even for relatively
well-separated chromophores because their respective sets of {\em
  remote} defects mutually overlap. In contrast, large-magnitude rare
jumps will sense only those defects found in the immediate vicinity
and thus will be uncorrelated. Thus according to
Ref.~\onlinecite{LSbinoculars}, two chromophores placed at a distance
$r \approx 200$~nm, at a temperature near 0.1 K, will decorrelate
after about one tenth of a millisecond.  The two-chromophore setup
allows one to separate out the concentration of the defects because it
introduces an additional lengthscale in the problem, viz., the
distance between the chromophores.  It would be interesting to
investigate whether such a setup could be used to differentiate
between the local and Yu-Leggett scenarios. Presumably, the non-local
nature of the compounds resonances implies the spectral jumps of two
remote chromophores will be correlated even on relatively long time
scales.

An entirely distinct avenue for tests of the molecular motions
underlying the cryogenic degrees of freedom is afforded by the
amorphous chalcogenides or any other materials that could host the
topological midgap electronic states. Just as in the context of
conjugated organic polymers,~\cite{RevModPhys.60.781} ENDOR
experiments have been proposed as a way to detect the mobile subset of
structural resonances.~\cite{ZLMicro2} In addition,
positron-annihilation lifetime spectroscopy (PALS) is a very sensitive
probe of defected configurations in glasses.  Note that the
traditional interpretation of PALS as a tool to probe
``micro-cavities'' does not seem to work well in the
chalcogenides.~\cite{NOVAK201729} One the other hand, positrons would
definitely interact the midgap states, which, recall, are {\em
  charged} in pristine samples; it is hoped that these experiments
will eventually enable one to determine the absolute, not just
relative concentration of positron scatterers. If directly confirmed,
the association between the structural resonances and the midgap
electronic states would indeed provide direct, strong support for the
RFOT-advanced, local scenario: In contrast with the massless phonons,
the wavefunction of an electron can extend, within a mobility gap, by
at most few nanometers. Based on this notion and many other
quantitative predictions of the RFOT theory,~\cite{LW_ARPC, L_AP} the
present author is convinced the physics underlying the cryogenic
anomalies is largely local but not single-particle; the extended
structural degrees of freedom are frozen out already above the glass
transition.

Last but not least, there has been recently a surge of interest in low
temperature anomalies in disordered films. In fact, TLS-like
excitations are deemed to be a key contributor to the loss of
coherence in Josephson junctions, which are an important candidate
system for quantum computing.~\cite{PhysRevLett.107.105504,
  PhysRevB.88.174202, 0953-8984-26-32-325401, Lisenfeld2017,
  PhysRevLett.116.167002} It is hoped that this contemporary and
fascinating application will revive interest in what many---though not
the present author!---would consider quite literally a cold case of
the contested relics of the liquid motions exhibited by cryogenic
disordered solids.

{\em Acknowledgements}: The author thanks his collaborators Peter
G. Wolynes, Andriy Zhugayevych, Dmytro Bevzenko, Pyotr Rabochiy, Jon
C. Golden, and Alexey Lukyanov. The author's work is supported by the
NSF Grant CHE-1465125 and the Welch Foundation Grant E-1765. He
gratefully acknowledges the use of the Maxwell/Opuntia Cluster and the
untiring support from the Center for Advanced Computing and Data
Systems at the University of Houston.  Partial support for this work
was provided by resources of the uHPC cluster managed by the
University of Houston and acquired through the NSF Award Number
ACI-1531814. Also gratefully acknowledged is the Texas Advanced
Computing Center (TACC) at The University of Texas at Austin for
providing HPC resources. URL: http://www.tacc.utexas.edu.


\bibliography{/Users/vas/Documents/tex/ACP/lowT}

\end{document}